# Multiscale Magnetic Correlations in $La_2Mn_{2-x}Ni_xO_6$: Role of Crystal Structure in Double Perovskites

**A. K. Bera*, K. S. Chikara, B. Saha[#], and S. M. Yusuf**

*Solid State Physic Division, Bhabha Atomic Research Centre, Mumbai-400085, India*
*Homi Bhabha National Institute, Anusaktinagar, Mumbai-400094, India*

**Mohd. Nasir and S. Sen**

*Department of Physics, Indian Institute of Technology Indore, Indore-453552, India*

**Contact author: akbera@barc.gov.in*
*[#] Present address: University of Missouri, Columbia, Missouri, 65211, United States.*

## Abstract

The magnetic correlations in double perovskites $La_2Mn_{2-x}Ni_xO_6$ ($x$=0.5, 0.75, 1.0, 1.25 and 1.5) have been systematically investigated across macroscopic, mesoscopic, and microscopic length scales using temperature-dependent bulk DC magnetization, neutron depolarization, and neutron powder diffraction measurements, respectivitly. The magnetic properties evolve from a long-range ferromagnetic (FM) order to a cluster ferromagnetic/spin-glass (FM/SG) behavior as the Ni concentration increases. This evolution is directly linked to changes in the crystal structure, transitioning from pure orthorhombic ($x$=0.5) to mixed orthorhombic and monoclinic ($x$=0.75–1.0), and eventually to mixed trigonal and monoclinic symmetries ($x$=1.25–1.5). Ni substitution enhances the magnetic ordering temperature from 170 K ($x$=0.5) to 280 K ($x$=1.0), but this is accompanied by a reduction in both magnetization and ordered magnetic moment. Beyond $x$=1.0, any long-range magnetic ordering is absent. Additionally, all compositions exhibit a reentrant spin-glass-like phase at low temperatures (below ~50 K). Neutron diffraction analysis confirms that long-range FM order occurs only in the orthorhombic phase, while the monoclinic and trigonal phases lack such magnetic ordering. The temperature-dependent magnetic correlations are closely connected to variations in crystal structural parameters, including lattice constants and unit cell volume. The electrical conductivity behavior, following the variable range hopping (VRH) model, highlights the role of multivalence Mn and Ni ions on the electrical properties. This study elucidates the microscopic mechanisms behind the tunable magnetic and electrical properties of $La_2Mn_{2-x}Ni_xO_6$, offering valuable insights for the design of advanced materials for spintronic applications.

# I. INTRODUCTION:

The growing demand for efficient, sustainable, and cost-effective spintronic devices has been driven by the increasing prevalence of plug-in technologies. While current spintronic devices rely on well-established commercial materials, advancements continue through the optimization of material morphology, synthesis methods, and device engineering for future spintronics technology [1-4]. However, these improvements have been incremental rather than transformative. The discovery of novel functional materials, particularly those enabled by selective chemical doping or substitution, is recognized as essential for achieving higher efficiency, stability, and functionality in future spintronic devices. Transition metal and rare-earth oxides are at the forefront of this research due to their intriguing crystal structures, magnetic properties, and electrical behaviors. These materials exhibit strong coupling between spin, charge, lattice, and orbital degrees of freedoms at the microscopic length scale, resulting in diverse magnetic and electrical phases under external stimuli, such as temperature, magnetic fields, electric fields, or pressure [5]. Such phenomena are commonly observed in materials with unique crystal structures, including spinel ($AB_2O_4$) [6-8], inverse spinel ($B(AB)O_4$) [9-11], Brownmillerite ($A_2B_2O_5$) [12-14], perovskite ($ABO_3$) [5,15-17], double perovskite ($A_2BB'O_6$) [5,18-20], etc. In these crystal structures, $A$, $B$, and $B'$ are transition metal or rare-earth cations with varying valence states and local environments. Among them, double perovskites with the general formula $A_2BB'O_6$ are of particular interest due to their crystal structural and compositional flexibility, enabling tunable magnetic and electrical properties suitable for advanced spintronic applications. These materials also offer opportunities to explore fundamental physics related to crystal structure, magnetic and electric properties at the microscopic level.

Research on double perovskites has increasingly focused on uncovering their remarkable magnetic properties, including magnetodielectric effects [21], magnetoresistance [22], superconductivity [23], and magnetocapacitance [24,25]. Of particular interest are ferromagnetic insulating double perovskites, which are well-suited for spintronic applications. These materials enable spin current (the flow of electron spin rather than charge) and its control, both of which are critical for device functionality. However, ferromagnetic behavior is often accompanied by metallicity [26], and ferromagnetic insulators remain relatively rare [27,28]. Despite this rarity, some double perovskite compounds combine ferromagnetic and insulating properties, making them highly attractive for spintronics and multiferroics [29,30]. The magnetic and electrical properties of double perovskites ($A_2BB'O_6$) are strongly influenced by the cation types at the $A$, $B$, and $B'$ sites [31-34]. For example, Mn-based double perovskites, such as $La_2MnFeO_6$ [35], $La_2MnCuO_6$ [35], $La_2MnRuO_6$ [36], $La_2MnNiO_6$ [35] and $La_2MnCoO_6$ [37], particularly when the transition metal ion at the $B'$ site varies, exhibit complex and rich magnetic and electric properties due to the multivalent states of Mn ($Mn^{2+,}$ $Mn^{3+}$, and $Mn^{4+}$) and $B'$ ion.

Among these, $La_2MnNiO_6$ (LNMO) stands out as a multifunctional material exhibiting ferromagnetism and semiconducting behavior near room temperature [38,39]. However, its crystallography is highly sensitive to synthesis conditions, and can stabilize in multiple crystal structural phases, viz., monoclinic ($P2_1/c$), rhombohedral ($R\bar{3}c$), orthorhombic (*Pbnm*), or mixed-phases [40]. Consequently, the crystal structural symmetry and site order/disorder of Mn and Ni ions directly impact the magnetic and transport behaviors. While the stoichiometric compound $La_2NiMnO_6$ with equal concentrations of Mn and Ni-ions stabilizes in mixed space groups $R\bar{3}c$ and orthorhombic *Pbnm* at room temperature and shows ferromagnetic-semiconductor behaviour, its parent end members are strikingly different: $LaMnO_3$ (*Pbnm*, an antiferromagnetic insulator) and $LaNiO_3$ ($R\bar{3}c$, a metallic paramagnet) [41,42]. Intermediate compositions $La_2Ni_xMn_{2-x}O_6$ ($0.5 \leq x \leq 1.5$) thus represent a rich platform to explore competing crystal structural phases and complex magnetic ground states [43-45]. Earlier reports by Blasco *et al*. [46] indicate that compounds with $x < 1.0$ favors orthorhombic space group *Pbnm*, while the compounds with $x > 1.0$ stabilize in $R\bar{3}c$, and at $x \approx 1.0$ very often a coexistence of both crystal structural phases occurs. Recent reports [46], indicate that the coexistence of the multiple crystal structural phases (including orthorhombic, trigonal as well as a new monoclinic phase) may exists over a wider range of compositions below and above $x$=1.0. Moreover, the true crystal structural symmetry for the Mn rich compounds (0.5<$x$>1.0) is debatable between site-ordered monoclinic $P2_1/c$ and site-disordered orthorhombic *Pbnm* phases [44,45]. Further, a huge variation (over two orders of magnitude) of the bulk magnetic properties of $La_2Ni_xMn_{2-x}O_6$ over the concentration range $0.5 \leq x \leq 1.5$ was reported [43-45]. In addition, the magnetic ground state for the intermediate compound with $x$=1 is completely different from that of the end members; $LaMnO_3$ having an ordered AFM state and LaNiO3 having a paramagnetic state [41,42]. Microscopic neutron scattering results available for the $x \leq 1.0$ compounds [46] report a single monoclinic $P2_1/c$ crystal structure at 1.5 K (in contrast to the coexistence of trigonal $R\bar{3}c$ and orthorhombic *Pbnm* phases at room temperature [46]), together with a reduced site-averaged ordered moment of 0.9 $\mu_B$/ion for $x$ = 1.0, and a single orthorhombic *Pbnm* crystal structure (at 1.5 K and room temperature) for the $x$ = 0.2 and 0.5 compounds, respectively. The microscopic origin of magnetism over the intermediate composition range ($0.5 \leq x \leq 1.5$) remains unresolved: does long-range ferromagnetism arise from the site-disordered trigonal $R\bar{3}c$ phase or site-disordered orthorhombic *Pbnm*-phase, or from the site-ordered monoclinic $P2_1/c$ phase, or from all of the three phases? More broadly, a coherent picture linking crystal structural phase evolution, microscopic magnetic correlations, and electronic behavior across the composition range $0.5 \leq x \leq 1.5$ is missing and needs a comprehensive exploration over the multiple (macroscopic, mesoscopic, and microscopic) length scales.

In this work, we have considered multiple compositions of $La_2Ni_xMn_{2-x}O_6$ ($x$ = 0.5, 0.75, 1.0, 1.25, 1.5) and present the results of a comprehensive multiscale investigation using macroscopic magnetization, mesoscopic neutron depolarization, microscopic neutron diffraction, and impedance spectroscopy. Systematic analyses have been performed to probe structure–property correlations. Our results reveal that as the Ni concentration increases over $0.5 \leq x \leq 1.5$, a crystal structural evolution occurs from the pure orthorhombic ($x$=0.5) to mixed orthorhombic and monoclinic ($x$=0.75–1.0), and eventually to mixed trigonal and monoclinic ($x$=1.25–1.5) symmetries. Subsequently, the magnetic properties evolve from a long-range ferromagnetic (FM) order to a cluster ferromagnetic/spin-glass (FM/SG), as confirmed by both microscopic and mesoscopic neutron scattering results, which is directly linked to the crystal structural evolution. Most importantly, the monoclinic phase appears in the intermediate composition range with a maximum at $x$ =1.0. Further, neutron diffraction analysis confirms that long-range FM order occurs only in the orthorhombic phase, while the monoclinic and trigonal phases lack such magnetic long-range ordering. As a result, no long-range magnetic ordering is present for $x > 1.0$ when the orthorhombic crystal structural phase is absent. Further results establish the interplay of cation concentration and their valence states in stabilizing magnetism as well as variations of conduction properties across the series; thereby providing crucial insights for tuning of magnetic and electrical properties of double perovskites. Therefore, the present work not only advances the fundamental understanding of the physical properties of double perovskites but also highlights their potential for spintronic applications, owing to their tunable magnetic and conduction properties.

## II. EXPERIMENTAL METHODS:

Powder samples of $La_2Mn_{2-x}Ni_xO_6$ ($x$ = 0.5, 0.75, 1.0, 1.25, and 1.5) were synthesized using the conventional sol-gel method [43-45]. The phase purity of the samples was verified using laboratory x-ray diffraction before conducting magnetization and neutron scattering studies.

Temperature-dependent magnetization [$M(T)$] measurements for all the compounds were recorded using a commercial vibrating sample magnetometer (Cryogenic Co. Ltd., UK) under a magnetic field of 250 Oe, over a temperature range of 2–300 K. Data were collected using three different measurement protocols: zero-field-cooled warming (ZFC), field-cooled cooling (FCC), and field-cooled warming (FCW).

Neutron powder diffraction (NPD) measurements were performed at the Dhruva Research Reactor, Bhabha Atomic Research Centre (BARC), Mumbai, India, using the PD-1 neutron powder diffractometer ($\lambda$ = 1.094 Å) equipped with three linear position-sensitive detectors covering a 2θ angular range of 5–70° (corresponding to a $Q$ range of 0.4–6.5 Å$^{-1}$) [33]. Data were collected at various temperatures between 5 and 300 K to study the evolution of crystal and magnetic structures. For these measurements, powder samples were packed inside of a cylindrical vanadium container (6 mm diameter, 50 mm length) and attached to the cold head of a helium-4 closed-cycle refrigerator (CCR). Vanadium containers were chosen due to their small coherent scattering cross-section for thermal neutrons, ensuring minimal contamination of diffraction peaks. Each sample was thermalized for ~30 minutes at the desired temperature before data collection. The NPD data were analyzed using the Rietveld refinement method implemented in the FullProf suite [47].

One-dimensional neutron depolarization measurements were conducted using the Polarized Neutron Spectrometer (PNS, $\lambda$=1.205 Å) at the Dhruva Research Reactor [48]. Samples were cooled from room temperature to 2 K using a helium-4 CCR under a 50 Oe magnetic field (necessary to maintain beam polarization at the sample position). The transmitted neutron beam polarization was measured as a function of temperature during the warming cycle, under the same applied field.

Electrical measurements were carried out using a PSM1735−NumetriQ impedance analyzer (Newtons4th Ltd., UK). Spectra were recorded with an AC electric field amplitude of 0.05 V at selected temperatures down to 90 K in vacuum. For these measurements, disc-shaped pellets (10 mm diameter) were prepared by cold press with pressure of ~10 ton and subsequently, annealed at 1173 K for 10 hours to increase density. The densities of pellets were achieved higher than 90% of the theoretical value. A uniform coating of silver paste was applied to both the sides of the sintered pellets and dried at 773 K before impedance data collection to remove any volatile components from the silver paste and to improve contact adhesion.

## III. RESULTS AND DISCUSSIONS:

### A. Crystal Structural Correlations in $La_2Mn_{2-x}Ni_xO_6$

The experimental room temperature neutron diffraction patterns for $La_2Mn_{2-x}Ni_xO_6$ ($x$ = 0.5, 0.75, 1.0, 1.25, and 1.5) along with the calculated patterns by the Rietveld refinement method are depicted in Figure 1(a-e). For the Mn rich $x$ = 0.5 compound, the Rietveld analysis confirms an orthorhombic crystal structure with the *Pbnm* space group (space group No. 62), which is in good agreement with the earlier report based on a x-ray diffraction study [43]. It is noteworthy that the orthorhombic crystal structure (with the space group *Pbnm*) was also reported for the $x$ = 0.0 compound $La_2Mn_2O_6$ (*i.e.*, Mn ion alone) [27]. However, the observed orthorhombic structure with space group *Pbnm* having the lattice parameters relation $b < c/\sqrt{2} < a$ [Fig. 2(e-g)], is distinguishable from the O′-orthorhombic ($c/\sqrt{2} < a <$ *b*) phase found in the orbitally ordered oxygen stoichiometric end compound $LaMnO_3$ [49,50]. With the increasing of the Ni concentration for the *x*=0.75 and 1.0 compounds, appearance of additional Bragg peak at ~13.96° (highlighted by yellow shading in Fig. 1 (f-j)), has been observed which cannot be accounted by the orthorhombic space group *Pbnm*. At the same time, a decrease in the intensity of the Bragg peaks [viz., (110)/(002) Bragg peaks at $2\theta$~ 16 deg., highlighted by green shading in Fig. 1 (f-h)] of the orthorhombic phase has been found. This reveals that the crystal structures of these compositions contain mixed phases. Our detailed Rietveld analyses of the neutron diffraction patterns reveal the presence of mixed orthorhombic (*Pbnm*) and monoclinic [space group $P2_1/c$ (No. 14); a subgroup of the *Pbnm*] phases.

The quantitative analyses show that with the increase in Ni concentration from *x*=0.5 to *x*=1.0, the phase fraction of the monoclinic phase with the space group $P2_1/c$ (space group No. 14) increases sharply at the cost of the orthorhombic (*Pbnm*) phase. The fractions of the orthorhombic and monoclinic phases are determined to be ~ 50.1±1.4 wt% and 49.9±1.4 wt% for *x* = 0.75 compound; and ~ 10.2±1.5 wt% and 89.8±1.5 wt% for *x* = 1.0 compound, respectively. With further increase in Ni concentration beyond *x* = 1.0, the characteristic Bragg peaks for the orthorhombic phase disappear completely and new additional Bragg peaks appear (e.g., the Bragg peak at $2\theta$ ~16.37°; highlighted by cyan shading in Fig. 1 (i-j)). The new set of the Bragg peaks are well indexed by another phase having trigonal crystal structure with space group *R*-3*c* (space group No. 167). It is found that with the increasing Ni concentration from *x*= 1.25 to 1.5, the intensities of the Bragg peaks from the trigonal phase increase continuously. On the other hand, the intensities of the Bragg peaks from the monoclinic phase decrease continuously, revealing a transformation of the monoclinic phase to the trigonal phase. The Rietveld analyses reveal that the fractions of the monoclinic and trigonal phases are ~ 60.6 ± 2.4 wt% and 38.8±2.4 wt% for the *x* = 1.25 compound; and ~ 30.1±3.1 wt% and 67.7±3.1 wt% for the *x* = 1.5 compound, respectively. However, a small amount of NiO impurity phase (with phase fractions of ~ 0.54±0.4 and 2.23± 0.1 wt%) has been identified in both *x* = 1.25 and 1.5 compounds. The appearance of the trigonal phase for the higher Ni concentration agrees with the earlier reports [45] as well as that reported for the end compound $La_2Ni_2O_6$ [41]. The Rietveld refined crystal structural parameters are given in Table I.

In summary, present in-depth crystal structural study involving careful analyses of the neutron diffraction patterns reveals the evolution of the crystal structural phases from pure orthorhombic phase to a mixed orthorhombic and monoclinic phase, then to a mixed trigonal and monoclinic phase with the increasing Ni concentration in $La_2Mn_{2-x}Ni_xO_6$ over *x* = 0.5 to 1.5. The evolution of the phase fractions is shown in Fig. 2(a). Although the lower and higher Ni-substituted compounds have the crystal structure similar to the end compounds *x*= 0 and 2, the appearance of the monoclinic ($P2_1/c$) phase for the intermediate Ni concentration range is interesting. The maximum phase fraction for the monoclinic ($P2_1/c$) phase is found in the compound with $x$=1.0. It was reported that the compound $La_2NiMnO_6$ ($x$ =1.0) is sensitive to the synthesis conditions having biphasic nature with a high temperature trigonal ($R$-3$c$) phase that transforms to a monoclinic ($P2_1/c$) phase or orthorhombic ($Pbnm$) phase at low temperatures [51]. It was reported that both the trigonal and monoclinic /orthorhombic phases coexist over a wide temperature range [40,52]. Similar effects of phase coexistence are also observed in other compound, viz., $BaTiO_3$ [53].

It may be pointed out here that both orthorhombic (having the lattice parameter ratio $a < c/\sqrt{2} < b$) and trigonal crystal structures contain a single Wyckoff position [4$b$ (1/2,0,0) for orthorhombic phase and 6$b$ (0,0,0) for trigonal phase] for the transition metal ions (*i.e.*, Ni and Mn ions) revealing a site disordered crystal structure [Figs. 3(a and c)]. On the other hand, the monoclinic crystal structure has two different Wyckoff positions [2$c$(0,1/2,0) and 2$d$(1/2,0,0) sites)] for the transition metal ions that provides a possibility for the cation ordering of Ni and Mn ions [Fig. 3(b)]. It is known that the crystal structure of double perovskites ($A_2BB'O_6$) may adopt different types of crystal symmetry, such as, random type, rock salt type or layered type depending on the $B$ and $B'$ cation arrangements. The driving force for the $B$ and $B'$ ordering is the ionic size and charge difference between both kinds of cations. It is remarkable that monoclinic phase for all the compounds shows a rock salt type crystal structure with an extraordinarily high pseudo-orthorhombic character as far as unit cell dimensions are concerned, where the value of the monoclinic angle $\beta$ is very close to 90° [Fig. 2(h)]. The significant different neutron scattering lengths for Ni (10.3 fm) and Mn (-3.73 fm) allow to quantitatively investigate the site occupancies of Ni and Mn ions over the two crystallographic sites [2$c$(0,1/2,0) and 2$d$(1/2,0,0) sites)] in the presently studied compounds. Our analyses of the neutron diffraction patterns reveal that the Mn and Ni cations are distributed over both the crystallographic sites 2$c$ and 2$d$ with site preferences. The estimated degree of site (cation) ordering and its variation with Ni concentration are shown in Fig. 2(b). The site (cation) ordering (o) is calculated following the relation $o=(n_{Ni(2c)}-x/2)/((\min(x, (2-x)))/2)*100$, where $n_{Ni(2c)}$ is the site occupancy of the Ni ions at the 2$c$ site of the Monoclinic phase with space group $P2_1/c$ (Table I). The Ni concentration dependent degree of site (cation) ordering curve follows a similar trend of that for the monoclinic phase fraction. It, therefore, may be concluded that the tendency of site ordering of the Mn and Ni ions is the origin of the intermediate monoclinic phase. It is very unusual to found that the maximum site ordering between Mn and Ni ions occurs for $x$=1.0 compound where the concentration of all the transition metal ($TM$) cations [$Mn^{3+}$(0.645 Å), $Mn^{4+}$(0.53 Å), $Ni^{3+}$ (0.56 Å), and $Ni^{2+}$ (0.69 Å)] were reported to be almost equal [Fig. 2(d)] [45]. This observation does not support that the origin of the observed site ordering of Mn and Ni ions is solely due to the ionic size effect, as average ionic radii for Mn-ions (~0.59 Å) and Ni-ions (~0.62 Å) are almost equal for the compound with $x$=1.0. In contrast, when the ionic size difference is at its maximum for the $x$= 0.5 (0.61 Å for Mn-ions and 0.69

Å for Ni-ions, respectively) and 1.5 (0.53 Å for Mn-ions and 0.603 Å for Ni-ions, respectively) compounds, the site ordering is found to be minimal. The concentrations of $Mn^{3+}$, $Mn^{4+}$, $Ni^{3+}$, and $Ni^{2+}$ cations at the *B* and *B′* sites have been shown in Fig. 2(d) as a function of Ni-concentrations. It is evident that the maximum site ordering occurs when the $Mn^{3+}$ and $Ni^{3+}$ concentrations are almost equal. The strongly Jahn-Teller low-spin $Ni^{3+}$ ($3d^7$, $t_{2g}^6e_g^1$) and high-spin $Mn^{3+}$ ($3d^4$, $t_{2g}^3e_g^1$) ions may be responsible for the observed site ordering, and the effect is maximum when their concentrations are almost equal [51]. This is further supported by the significantly different average *TM*-O bond lengths for the two *B* and *B′* sites in the monoclinic phase [Fig. 2(c)]. The variation of lattice constants and unit cell volume with the Ni concentration are shown in Fig. 2(f-g). The lattice parameters and unit cell volume show an overall decrease with the increasing Ni-concentration which is in line with the decrease of the average ionic radius of the *B*/*B′* site from 0.65 Å for $x$=0.5 to 0.57 Å for $x$ =1.5.

### B. Macroscopic Magnetic Correlations in $La_2Mn_{2-x}Ni_xO_6$:

The temperature dependent bulk dc magnetization curves for $La_2Mn_{2-x}Ni_xO6$ ($x$= 0.5, 0.75, 1.0, 1.25 and 1.5), measured under a 250 Oe magnetic field using three different measurement protocols: ZFC, FCC, and FCW conditions are shown in Figure 4 (a-e). For the $x$=0.5 compound (i.e., the Mn rich compound), the temperature dependent magnetization curves reveal a FM transition at $T_c$~ 162 K with an additional anomaly at $T^*$~ 50 K which was reported to be due to a reentrance cluster spin-glass transition [45]. The FM ordering is in sharp contrast to the AFM ordering in the pure Mn-based compound $LaMnO_3$ with only the $Mn^{3+}$ ions [54]. The substitution of $Ni^{2+}$ introduces a mixed valence Mn-ions ($Mn^{3+}$ and $Mn^{4+}$) that leads to the introduction of FM double exchange interactions along with FM super exchange interactions (between $Ni^{2+}$–$Mn^{4+}$, $Ni^{2+}$–$Mn^{3+}$, and low-spin $Ni^{3+}$–high-spin $Mn^{3+}$ pairs), and AFM super exchange interactions (between $Ni^{2+}$–$Ni^{2+}$, $Ni^{3+}$–$Ni^{3+}$, $Mn^{3+}$–$Mn^{3+}$, and $Mn^{4+}$–$Mn^{4+}$ pairs). The magnetic properties of $La_2Mn_{1.5}Ni_{0.5}O_6$ ($x$=0.5) are thus dictated by the competing AFM and FM exchange interactions. Such competing interactions are believed to be the cause for the appearance of the low temperature reentrant cluster spin-glass transition below the $T^*$~ 50 K. With the increasing Ni concentration for $x$= 0.75 compound, a sharp enhancement of the FM ordering temperature to $T_c$ ~ 250 K is evident. Additional anomalies are observed at $T^\#$~ 190 and $T^*$ ~ 35 K which are linked to a FM ordering due to vibronic super exchange interaction between the low-spin $Ni^{3+}$ ($3d^7$, $t_{2g}^6e_g^1$) and $Mn^{3+}$ ($3d^4$, $t_{2g}^3$,$e_g^1$) pairs [55,56] and the spin-glass transitions, respectively. For the $x$=1 compound, the FM transition is further enhanced to $T_c$ ~ 290 K. Interestingly, only a single low temperature anomaly is observed at $T^{**}$~ 100 K. With further increases in the Ni concentration for $x$=1.25 compound, the FM transition temperature decreases to $T_c$ ~ 273 K and the low temperature anomalies reappear at $T^\#$~ 235

K and at $T$*~ 40 K, respectively. For the $x$ =1.5 compound, the FM transition temperature remains almost constant at $T_c$ ~ 270 K. The intermediate temperature anomaly at $T^{\#}$ disappears. The low temperature anomaly due to the spin-glass ordering shifts slightly to a higher temperature $T$*~ 44 K.

Here, we would like to mention that bifurcations have been observed between ZFC and FCC/FCW curves for all the compounds right from the high temperature FM transition temperature $T_c$. Such bifurcations may indicate that the magnetic ordered state either deviates from a pure ferromagnetic state or involves kinetics of magnetic domain effects [57]. The presence of competing FM and AFM exchange interactions may result a deviation from the pure FM state. The true microscopic natures of the magnetic phases were unknown and have been investigated in the present study employing neutron scattering techniques (discuss below). It may also be noticed that the absolute value of the magnetization decreases monotonically with the increasing Ni concentration. The magnetization value decreases by almost two orders of magnitude for the $x$=1.5 compound as compared to that for the $x$= 0.5 compound (for comparison see the values at the lowest measured temperature at 2 K). Such a huge reduction in the magnetization value from $x$ = 0.5 to 1.5 compound cannot be accounted by the conversion of the higher spin value of $Mn^{3+}$ ($S$=2) and $Ni^{2+}$ ($S$=1) ions to the lower spin-value of $Mn^{4+}$ ($S$=3/2) and $Ni^{3+}$ ($S$=1/2) ions with the increasing of the Ni concentration. The reduced value of magnetization, thus, must be related to the inherent change in the spin configurations or correlations that appear as a result of the competing FM and AFM exchange interactions. The nature (FM or AFM) of exchange interaction is strongly related to the combination of the nearest neighbor magnetic pairs i.e., $Ni^{2+}$–$Mn^{4+}$ (FM), $Ni^{2+}$–$Mn^{3+}$(FM), and low-spin $Ni^{3+}$–high-spin $Mn^{3+}$ (FM)), $Ni^{2+}$–$Ni^{2+}$(AFM), $Ni^{3+}$–$Ni^{3+}$(AFM), $Mn^{3+}$–$Mn^{3+}$(AFM), and $Mn^{4+}$–$Mn^{4+}$ (AFM).

### C. Mesoscopic Magnetic Correlations in $La_2Mn_{2-x}Ni_xO_6$:

To explore the nature of the magnetic state at the mesoscopic length scale (domain magnetization), we performed one-dimensional neutron depolarization analysis measurements on all compositions ($x$ = 0.5–1.5). Neutron depolarization analysis is a powerful technique for studying magnetic correlations at mesoscopic length (domain magnetization) scales in ferromagnetically/ferrimagnetically ordered compounds or cluster spin-glass systems. The final neutron polarization ($P_f$) after passing through randomly oriented magnetic domains in a ferromagnetic sample is described by the equation [58]:

$$P_f = P_i \, exp\left\{-\alpha \frac{d}{\delta} \langle \phi_\delta \rangle^2\right\}$$

where $P_i$ is incident neutron beam polarization, $\alpha$ is a dimensionless parameter, $d$ is sample thickness, and $\langle \phi_\delta \rangle$ is average precession angle per magnetic domain.

Figures 4(g–l) display the temperature-dependent transmitted neutron polarization ($P_f/P_i$) for $La_2Mn_{2-}$

${}_{x}Ni_{x}O_6$ ($x$ = 0.5, 0.75, 1.0, 1.25, and 1.5) measured under a 50 Oe magnetic field using the FCW protocol. To compare with bulk magnetization data, we plotted the FCW magnetization curves alongside the depolarization curves for each compound.

For the $x$ = 0.5 Mn-rich compound, a rapid decrease in neutron beam polarization below the first transition temperature $T_c$ signals the onset of ferromagnetic (FM) ordering, i.e., the formation of FM domains. As the temperature decreases further, the depolarization curve exhibits a Brillouin-type behavior, approaching saturation below approximately 100 K. Below ~60 K, an additional contribution to neutron beam depolarization becomes apparent (reaching approximately 4-5% at 4 K) [Figs. 4(g) and 4(l)]. The onset of this additional depolarization is closely aligned with the temperature $T^*$=50 K, which was identified as the reentrant cluster spin-glass transition temperature from susceptibility studies [45]. This additional depolarization component at low temperatures suggests the presence of distinct physical phases for the clusters spin glass state and FM state, indicating an electronic phase separation in the compound.

As Ni concentration increases, a significant reduction in overall depolarization values is observed, indicating a decrease in domain magnetization, i.e., ferromagnetic order at the domain level. For compounds with $x\geq1.25$, depolarization values become negligibly small. It is noteworthy that all low-temperature anomalies observed in the dc-magnetization curves are also evident in the temperature-dependent depolarization curves (highlighted by vertical lines), confirming their origin at the magnetic domain level.

### D. Microscopic Magnetic Correlations in $La_2Mn_{2-x}Ni_xO_6$:

To investigate the microscopic spin-spin correlations and determine the magnetic ground state of the studied compounds, we carried out temperature-dependent neutron powder diffraction (NPD) measurements down to 5 K. The temperature-dependent neutron diffraction patterns for $La_2Mn_{2-x}Ni_xO_6$ ($x$ = 0.5, 0.75, 1.0, 1.25 and 1.5) are presented in Fig. 5 (a-e). For all compositions, the neutron diffraction patterns remain unchanged down to 5 K, except an enhancement of some of the low angle Bragg peaks for few compounds due to magnetic ordering. This confirms an absence of any structural phase transitions over the entire temperature range 5-300 K. For the $x$ = 0.5 compound (i.e., Mn-rich), an increase in intensity of the Bragg peaks around $2\theta$ = 16° and 23° below $T_c$ = 162 K confirms the onset of long-range ferromagnetic (FM) ordering.

As Ni concentration increases, the presence of magnetic Bragg peak intensities is observed for the $x$ = 0.75 and $x$ = 1.0 compounds below their respective $T_c$ values. The consistent nature of the magnetic diffraction patterns for the compounds $x$ =0.5, 0.75 and 1.0 suggests that the magnetic structure remains unchanged with increasing Ni concentration. However, there is a noticeable decrease in the intensity of the magnetic Bragg peaks at 5 K, accompanied by an increase in peak width, indicating a reduction in the magnetic correlation length and, consequently, a decrease in magnetic domain size [Fig. 5(f)], in agreement with the neutron depolarization results [Fig. 4(g-l)]. For the Ni-rich compounds with $x \geq 1.25$ (i.e, $x$ = 1.25 and 1.5), no magnetic Bragg peaks are observed down to 5 K, suggesting an absence of long-range magnetic ordering or the presence of weak magnetic peak (corresponding the magnetic moment < 0.5$\mu_B$) magnetic moments below the resolution limit of the neutron diffractometer. The monotonous reduction of magnetic Bragg peak intensities and their complete disappearance for $x$>1. 0 are consistent with the previously discussed bulk magnetization and neutron depolarization results [Fig. 4].

The experimentally measured (at 5 K) and Rietveld-fitted neutron diffraction patterns for all five samples ($x$ = 0.5, 0.75, 1.0, 1.25, and 1.5) are shown in Fig. 6(a-e). The observed magnetic Bragg peaks could be indexed with the FM propagation vector $k$=(000) with respect to the *Pbnm* nuclear unit cell. The magnetic peaks could not be indexed using the $P2_1/c$ crystal unit cell, confirming that the FM contribution arises solely from the *Pbnm* phase. Therefore, the decrease in bulk ferromagnetic strength is attributed to the phase fraction (wt%) of the *Pbnm* phase, which decreases monotonically with increasing Ni concentration [Fig. 2] and vanishes for $x$ >1.0. A good agreement between the experimental and calculated patterns is obtained through either two-phase refinement (nuclear and magnetic phases of *Pbnm* for the compound $x$ = 0.5) or three-phase refinement (nuclear and magnetic phases of *Pbnm*, and the nuclear phase of $P2_1/c$ for the compounds $x$ = 0.75 and 1.0) [Fig. 6]. These analyses reveal that the magnetic moments are aligned along the crystallographic $c$-axis [Fig. 6(f)] in a collinear FM structure. Due to the mixed occupation of Mn and Ni ions at the $B/B'$ crystallographic sites in the *Pbnm* phase, it is not possible to determine the individual ordered magnetic moments for Mn and Ni ions. However, the site-averaged ordered magnetic moments are estimated from the neutron diffraction data for all ferromagnetically ordered compounds. For the $x$ = 0.5 compound, the ordered magnetic moment is determined to be 2.76(4) $\mu_B$/ion at 5 K, which is in good agreement with previously reported value [59]. With increasing Ni concentration, the site-averaged ordered moment decreases monotonously, reaching approximately 1.85 $\mu_B$/ion at 5 K for the $x$ = 1.0 compound [Fig. 6(g)]. At the same time, the width of the magnetic Bragg peaks increases, indicating a reduction in the FM correlation length and, consequently, a decrease in FM domain size.

The temperature-dependent refined ordered magnetic moment values for $B/B'$ site are shown in Figs. 7(a-c) for the $x$ = 0.5, 0.75, and 1.0 compounds, respectively. The curves exhibit finite moment values below the corresponding FM ordering temperatures ($T_C$) and additional anomaly at $T^{\#}$ corresponding to the FM ordering due to vibronic super exchange interaction between the low-spin $Ni^{3+}$ ($3d^7$, $t_{2g}^6e_g^1$) and $Mn^{3+}$ ($3d^4$, $t_{2g}^3$,$e_g^1$) pairs. The temperature variations of the lattice parameters and unit cell volume for the $x$ = 0.5 compound are shown in Figs. 7(d-g). The curves display subtle anomalies at $T_C$, suggesting a possible magneto-structural coupling.

## E. Evoluation of the Electrical Propeties in $La_2Mn_{2-x}Ni_xO_6$:

The electrical properties of the double perovskite compounds are highly influnced by the relative occupancies of the transition metal ions at the $B$ and $B'$ sites which can be used for the tuning of the electrical behaviour of the studied compounds to achieve optimum properties. The temperature dependent resistivity values over the temperature range of 90–300 K are shown in Fig. 8(a) for the compounds $x$ = 0.5, 0.75, and 1.0. For all the compounds (0.5 ≤ $x$ ≤ 1.0), the temperature dependent resistivity curves reveal a semiconducting behaviour with an increase in the resistivity value with the increasing Ni concentration. The values of the resistivity for $x$ = 0.5 and 1.0 are in good agreement with that reported ealier [60,61]. Whereas, beyond the $x$ =1.0 (Ni rich compounds with $x$ =1.25 and 1.5), the samples become highly conductive and the resistivity values become too small (for all the temperature range) to be measured using our instrument. The highly conductive behaviour for the compound with $x$ =1.5 was also reported in litrature [44] and the reported curve is included in Fig. 8(a) for a comparison. The highly conducting behaviour of the Ni-rich compounds is consistent with the end compound $LaNiO_3$, which has a metallic nature. The change in the resistivity is attributed to the mixed valence states of Ni and Mn as their ratio changes upon variation of $x$-valus.

In the literature [61,62], the various models, such as, Arrhenius-type, nearest-neighbor hopping (NNH), and variable range hopping (VRH) were used to explain the conduction mechanism in the double perovskite materials. The Arrhenius plot (ln$\rho$ vs 1000/$T$) of the resistivity curves are shown in Fig. 8(b). It may be noted that the cuves are linear for the low temperature range over ~ 90-150 K following the term: $\rho_{dc} = \rho_0 exp(E_{dc}/k_BT)$ and then show a convex type nature for higher temperatures. Here, $\rho_0$ is the prefactor and $E_{dc}$ is the activation energy. The variation of the activation energy with the concentration of Ni ions is shown in the inset of Fig. 8(b). The activation energy is found to be ~ 1.2 eV and a slight increase has been found with the increasing $x$-value. The deviation from the linear behaviour in the higher temperature range (> ~150 K) indicates that better models are required to determine the microscopic mechanism of the resistivity. Polaronic models [43,58] were extensively used in literature to interpreate the microscopic mechanisms of electrical conductivity in double perovskite oxide materials. In the small polaron model there could be two distinct types hoping processes i.e., nearest-neighbor hopping (NNH) and variable range hopping (VRH). For the NNH model, the hopping occurs between two neighbouring sites (in an ordered crystal lattice) and the mobility temperature dependence is presumed. On the other hand, the VRH model considers hopping of distances distributed over space (in a slightly or locally disordered lattice) due to random distribution of non-uniform potential wells resulting from structural disorder.

In the framework of the NNH small polaron adiabatic model, temperature dependent resistivity curve follows the relation $\rho_{dc}(T) = \rho_0 T exp(E_{dc}/k_B T)$. We have shown the $ln(\rho/T)$ vs. 1000/$T$ plots for all the three compounds in Fig. 8(c). However, all the curves show convex type nature, especially for the high temperature range, indicating that the NNH model is not valid for the present compounds. Next, the VRH small polaron model was checked. According to this model, the resistivity varies with temperature following the equation,

$$\rho(T) = \rho_0 \exp\left(\frac{T_0}{T}\right)^{\frac{1}{4}}$$

where, $\rho_0$ is the prefactor, and $T_0$ is the hopping temperature, which is related to the hopping probability and a measure of local disorder. The temperature dependent resistivity, plot following the VRH model [Fig. 8(d)], show a linear behaviour for the higher temperature range (150-300 K). The linear fitting of the curves over 150–300 K yields $T_0$ value in the range 10-30×$10^7$ K. The fitting parameters and the agreement factor (adjusted $R^2$) for both the NNH and VRH models are included in Table II. The variation of the $T_0$ parameter with the Ni-concentration $x$ is shown in the inset of Fig. 8(d). The $T_0$ value is in agreement with those observed for similar double perovskites [43]. The presence of VRH conduction phenomena hinted at *B*-site lattice disorder which has been found from the neutron diffraction results and discussed in the crystal structural section. The VRH mediated electronic conduction was also reported for the related compounds $La_2MnB'O_6$ ($B'$ = Fe, Ni, and Zn) [35].

## IV. DISCUSSION:

As outlined in the introduction and result sections, the physical properties of double perovskites $A_2BB'O_6$ are highly sensitive to synthesis conditions and to the relative concentrations of transition-metal ions at the *B* and $B'$ sites. In this context, our systematic study of $La_2Mn_{2-x}Ni_xO_6$ ($x$ =0.5,0.75,1.0,1.25, and 1.5) provides a benchmark for magneto–structural–electrical correlations and serves as a reference for comparison with other La–Mn–based double perovskites, such as $La_2Mn_{2-x}B'_xO_6$ ($B'$ = Ni, Co, Fe, Cr, Ti, Ru, Cu)[35] [63]. Present neutron diffraction results reveal a clear crystal structural evolution in $La_2Mn_{2-x}Ni_xO_6$: a pure orthorhombic *Pbnm* phase in the Mn-rich regime ($x \leq 0.5$), a mixed orthorhombic–monoclinic (*Pbnm* + $P2_1/c$) phase at intermediate compositions ($0.75 \leq x \leq 1.0$), and a monoclinic–trigonal ($P2_1/c$ + $R$-3$c$) mixture in the Ni-rich region ($x \geq 1.25$). Similar crystal structural transformations have been reported in the related $La_2Mn(Co,Ru)O_6$ system [63], corresponding to the effective composition range $0.5 \le x \le 1.0$ for the present series $La_2Mn_{2-x}Ni_xO_6$. In $La_2Mn(Co_{1-x}Ru_x)O_6$, Ru substitution at the Co ($B'$) site drives a transition from a disordered orthorhombic *Pnma* phase to an ordered monoclinic $P2_1/n$ phase at $x \approx 0.3$. However, other studies on $La_2Mn(Co_xM_{1-x})O_6$ ($M$ = Fe, Cr, Zn, Ni) [64-66] suggest that all substituted compounds stabilize in the monoclinic $P2_1/n$ structure. In contrast, Cu substitution in $La_2MnCo_{1-x}Cu_xO_6$ stabilizes the orthorhombic *Pbnm* phase [35]. Remarkably, unlike these related systems, $La_2Mn_{2-x}Ni_xO_6$ exhibits a robust coexistence of orthorhombic–monoclinic (*Pbnm* + $P2_1/c$) phases over a wide composition range.

Now we focus on the effect of relative concentrations of the cations at the $B$ and $B'$ sites and their relative valence states on the crystal structure and magnetic properties of double perovskite compound $La_2Mn_{2-x}Ni_xO_6$. From the results of the present study and literature, strong effects of Mn:Ni ratio on the crystal structural, magnetic and electrical properties have been evident for $La_2Mn_{2-x}Ni_xO_6$. The end compounds of the series, namely $LaMnO_3$ and $LaNiO_3$, have the Mn and Ni in the +3 valence state, respectivley. The substitution of the Ni at Mn site in the $LaMnO_3$ compound creates the charge imbalance. To compenste the charge imbalance and maintain the charge neutrality, both Mn and Ni ions exist in multivalence states i.e., $Mn^{3+}$, $Mn^{4+}$, $Ni^{2+}$, and $Ni^{3+}$ according to the following relation,

$$i.Mn^{3+} + k.Ni^{3+} \leftrightharpoons l.Ni^{2+} + j.Mn^{4+}$$

where $i$, $j$, $k$, and $l$ are weight fractions of each valence state in the single formula unit (with constraint such that $i + j = 2 - x$ and $k + l = x$) of respective compound. Thus, the equilibrium may be driven either to the left or to the right in the above equation depending on several factors. For the studied series of compounds $La_2Mn_{2-x}Ni_xO_6$ with $x$ = 0.5, 0.75, 1.0, 1.25, and 1.5, the ionic formula reported from XPS analysis are: $La^{3+}{}_2(Ni^{2+}{}_{0.5}Ni^{3+}{}_{0.0})(Mn3^{+}{}_{0.989}Mn^{4+}{}_{0.511})O^{2-}{}_6$, $La^{3+}{}_2(Ni^{2+}{}_{0.46}Ni^{3+}{}_{0.29})(Mn^{3+}{}_{0.82}Mn^{4+}{}_{0.43})O^{2-}{}_6$, $La^{3+}{}_2(Ni^{2+}{}_{0.52}Ni^{3+}{}_{0.48})(Mn^{3+}{}_{0.48}Mn^{4+}{}_{0.52})O^{2-}{}_6$, $La^{3+}{}_2(Ni^{2+}{}_{0.57}Ni^{3+}{}_{0.68})(Mn^{3+}{}_{0.36}Mn^{4+}{}_{0.39})O^{2-}{}_6$, and $La^{3+}{}_2(Ni^{2+}{}_{0.5}\,Ni^{3+}{}_{1.0})(Mn^{3+}{}_{0.0}Mn^{4+}{}_{0.5})O^{2-}{}_6$, respectivlty [44,45,60]. Here, we would like to mention that as the Ni concentration increases, the ratio of $Ni^{3+}/Ni^{2+}$ increases and $Mn^{3+}/Mn^{4+}$ decreases [Fig. 2(d)]. This trend is inline with the decreasing ferromagnetic component with increasing Ni concentration. Here, we point out that, for $x$ =0.5 compound, there is an absance of $Ni^{3+}$ and the compound ( $x$ = 0.5) forms into a single crystal structural phase, well characterized by the ortherhombic *Pbnm* space group. Further increase of the Ni concetration, the $Ni^{3+}$ gets introduced in the materials which leads to a reduction of the $Mn^{3+}$ to $Ni^{3+}$ ratio [Fig. 2(d)], and results into a new structural monoclinic phase ( $P2_1/c$) for $x$ =0.75. Further increase of the Ni concetration, the $Mn^{3+}$ to $Ni^{3+}$ ratio becomes unity at $x$ =1.0 which is inline with the increase of fraction of the monoclinic phase with a maximum at $x$ =1.0. Increase of the Ni concentration beyond the value $x$ =1.0, the number of $Ni^{3+}$ ions becomes higher than the $Mn^{3+}$ ions and a decrease of the monoclinc phase $P2_1/c$ fraction is evident. At the same time, the orthorhombic phase completely disappers for $x > 1.0$ and a trigonal phase ($R$-3$c$) appears for the Ni reach compounds ($x$ >

1.0). The end compound $LaNiO_3$ (i.e., $x$=2.0) is reported to have only the trigonal ($R$-3$c$) crystal structure. It is discussed earlier that as the $B$ and $B'$ sites contain ions having almost similar ionic radius, and the observed anti-site ordering may be due to the Jahn-Teller distortions containg the low-spin $Ni^{3+}$ ($3d^7$, $t_{2g}^6e_g^1$) and high-spin $Mn^{3+}$ ($3d^4$, $t_{2g}^3e_g^1$) ions [51]. The anti-site ordering is found to be maximum when the $Mn^{3+}$ to $Ni^{3+}$ ratio is unity (i.e., for the compound having $x$ = 1.0).

We now address the symmetry dependence of the magnetic correlations. Our neutron scattering results reveal that long-range ferromagnetic order is stabilized only in the orthorhombic *Pbnm* phase of $La_2Mn_{2-x}Ni_xO_6$, while no long-range magnetic ordering is observed in the monoclinic $P2_1/c$ or trigonal R-3c phases. Previous work by Blasco *et al.* [46,67] reported long-range magnetic order for all compositions with $x$≤1.0. Although the $x$=0.2 and 0.5 compounds were identified as orthorhombic, the $x$= 1.0 compound was assigned a monoclinic ($P2_1/n$) symmetry at low temperature, and the magnetic ordering was analyzed within the monoclinic framework. We note, however, that their reported magnetic pattern closely resembles our magnetic pattern [Fig. 5]. Our systematic diffraction investigation on the $x$=0.5, 0.75, 1.0, 1.25, and 1.5 compounds establish a direct correlation between the orthorhombic phase fraction and magnetic peak intensity: the latter decreases monotonically with increasing Ni concentration, reaching a minimum at $x$=1.0 as like the orthorhombic phase fraction. In contrast, the composition dependence monoclinic phase fraction is opposite to the variation of the magnetic peak intensity. The magnetic peak intensity reaches to a minimum at $x$=1.0 where the monoclinic phase dominates (~ 89.8±1.5 wt%), and vanishing entirely for $x$=1.25, where the monoclinic fraction exceeds 50%. These results demonstrate unambiguously that long-range ferromagnetic order is intrinsic to only the orthorhombic Pbnm phase. This clarification, absent from previous reports, represents the first such identification in double perovskites and suggests that similar investigations on $La_2Mn_{2-x}B'_xO_6$ ($B'$ = Ni, Co, Fe, Cr, Ti, Ru) are warranted to test the universality of this behavior.

Now we discuss the origin of different magnetic phase transitions as well as shed light on the observed decrease in the ferromagnetism with increasing the Ni-concentration in $La_2Mn_{2-x}Ni_xO_6$. As we discussed earlier, the magnetic properties of the double perovskite compounds highly susceptible to the valence state of the cations at $A$, $B$ and $B'$ sites. For the present series of compounds, the $B$ and $B'$ sites have been occupied by the megnetoactive cations (Mn and Ni), whereas, $A$ site has been occupid by nonmagnetic cations (La). As we have discussed above, the ionic formula for all the compounds reveals the presence of mixed valence states of Mn and Ni (*i.e.*, $Mn^{4+}$, $Mn^{3+}$, $Ni^{2+}$, and $Ni^{3+}$), subsequently, mixed spin values. The concentration of each of the ions (*i.e.*, $Mn^{4+}$, $Mn^{3+}$, $Ni^{2+}$, and $Ni^{3+}$) vary with the Ni concentrations resulting into the possibilities of different type of nearest neighborus ion-pairs and subsequently different types (FM and/or AFM) of exchange interactions. Further, the microscopic origin of the different magnetic properties of these compounds can be understood through the competeing FM/AFM superexchange and FM double exchnage interactions among the magnetically active transition metal ions mediated by the oxygen ligands [55]. According to the Goodenough-Kanamori rule [50,56], the superexchnage interection between two magnetic ions (mediated through the non-magneitc ions) strongly depends on the metal-nonmagnetic-metal bond angle. For example, the superexchange interaction between two orthogonal orbitals is favorable for parallel allignment of spins (ferromagnetic), whereas the superexchange interaction is favorable for antiparallel allignment of spins (antiferomagnetic) if magnetic orbitals have the same symmetry. Here, we whould like to mention that for each compound, the presence of multiple valance states of Mn and Ni-ions results into competing FM and AFM exchange interactions. AFM superexchnage interactions occur for the pathways (i) $Ni^{2+}$-O-$Ni^{2+}$, (ii) $Ni^{2+}$-O-$Ni^{3+}$, (iii) $Ni^{3+}$-O-$Ni^{3+}$ , (iv) $Mn^{3+}$- O- $Mn^{3+}$ and (v) $Mn^{4+}$- O- $Mn^{4+}$; on the other hand, FM superexchange interactions occur for (vi) $Ni^{2+}$-O-$Mn^{3+}$, (vii) $Ni^{2+}$-O-$Mn^{4+}$, (viii) $Ni^{3+}$-O-$Mn^{4+}$, and (ix) $Ni^{3+}$-O-$Mn^{3+}$ pathways; as well as FM double exchange interactions occur for (x) $Mn^{3+}$-O-$Mn^{4+}$ pathways. For the compound with $x$ = 0.5 (without having the $Ni^{3+}$ ions), the FM superexchange interactions for $Ni^{2+}$-O-$Mn^{3+}$ and $Ni^{2+}$-O-$Mn^{4+}$ as well as FM double exchange interactions for $Mn^{3+}$-O-$Mn^{4+}$are dominating in Mn reach compounds leading to the observed long-range FM ground state. With the increasing Ni substitution, the appearance of $Ni^{3+}$ ions results into introduction of additional four possible exchange interactions for the pathways $Ni^{2+}$-O-$Ni^{3+}$ (AFM), $Ni^{3+}$-O-$Ni^{3+}$ (AFM), $Ni^{3+}$-O-$Mn^{3+}$ (FM), and $Ni^{3+}$-O-$Mn^{4+}$(FM). The introduction of these additional exchange interactions increases bond randomness and resulting into a loss of long-range ferromagnetic ordering, leading to a short-range (cluster) FM correlations. However, the FM interaction strength grows with substitution as the concentration of the $Ni^{3+}$ rises which increases the possibility of the FM $Ni^{3+}$-O-$Mn^{4+}$ exchange interactions, leading to an enhancement in the transition temperature of ferromagnetic correlations [Fig. 6(h)]. The reduction of the ordered moment values with the increasing $x$ value is related to the introduction of the $Ni^{3+}$ ions with spin value of ½ at the cost of $Mn^{3+}$ ions carrying a spin value of either $S$=2 for high spin configuration or $S$ = 1 for the low-spin configuration. Such a reduction of the site averaged spin value is accountable for the observed decrease in the site averaged ordered moment value from 2.76 $\mu_B$/ion (for $x$ =0.5) to 1.85 $\mu_B$/ion (for $x$=1.0). The absence of long-range magnetic ordering in the intermediate monoclinic $P2_1/c$ phase is due to partial cation disorder, competing magnetic interactions, structural distortion, and magnetic frustration. These factors collectively disrupt the conditions necessary for coherent magnetic alignment, resulting in a state dominated by short-range correlations or dynamic spin fluctuations. The short-range correlations are evident from the neutron depolarization results. On the other hand, the absence of long-range magnetic ordering in the Ni-rich $R$-3$c$ phase is primarily due to its metallic nature, strong Ni-O hybridization, itinerant electron behavior, and the lack of structural distortions as reported for the end compound $LaNiO_3$. These factors collectively inhibit the formation of static magnetic moments and long-range spin-spin correlations.

Electrical resistivity in $La_2Mn_{2-x}Ni_xO_6$ exhibits semiconducting behavior across the entire series ($0.5 \leq x \leq 1.5$). With the increasing Ni content, conductivity value first decreases up to $x$ =1.0 and then enhances with further increase of the Ni concentration. The values of the resistivity for $x$ = 0.5 and 1.0 are in good agreement with that reported ealier [60,61]. For the related compounds $La_2MnCo_{1-x}Ni_xO_6$ ($x$ = 0, 0.1 and 0.5), it was reported that conductivity value decrease with the Ni substitution on the Co site [68] (as found in the present study) which was attributed to reduced carrier density. These parallels suggest that in $La_2Mn_{2-x}Ni_xO_6$, Ni substitution modifies charge-carrier localization (in addition to magnetic interactions) and hopping dynamics, reinforcing the intimate coupling between crystal structure, electronic transport, and magnetism. The transport mechanism for the presently studied compounds is governed by variable-range hopping (VRH) mediated by mixed-valence Mn/Ni ions, consistent with the reported related compounds $La_2MnB'O_6$ ($B'$ = Fe, Ni, and Zn) [35]. The temperature dependence of resistivity of $La_2MnCo_{1-x}Ni_xO_6$ systems was well described by VRH conduction method over 194–300 K for $x$ = 0.1 and over 160–310 K for $x$ = 0.5 [68]. In short, the present results intimate a tunable electronic transport properties coupling with crystal structure, and magnetism of double perovskites $La_2Mn_{2-x}Ni_xO_6$.

## SUMMARY AND CONCLUSION:

We have carried out a comprehensive investigation of the magnetic properties of the double perovskite series $La_2Mn_{2-x}Ni_xO_6$ ($x$ = 0.5, 0.75, 1.0, 1.25, and 1.5) across macroscopic, mesoscopic, and microscopic length scales using DC magnetization, neutron depolarization, and neutron powder diffraction measurements. Rietveld refinement of neutron diffraction data reveals a systematic evolution of the crystal structure: a single orthorhombic phase (*Pbnm*) for $x$ = 0.5, coexisting orthorhombic and monoclinic ($P2_1/c$) phases for $x$ = 0.75 and 1.0, and a mixed monoclinic–trigonal ($P2_1/c$ + $R\text{-}3c$) symmetry for $x$ = 1.25 and 1.5. The fractional contributions of each phase were quantitatively determined, showing that the $P2_1/c$ phase appears predominantly in the intermediate range and reaches its maximum at $x$=1.0, before decreasing with further Ni substitution. Through comprehensive neutron diffraction studies, we have confirmed that the orthorhombic phase exclusively exhibits a long-range ferromagnetic (FM) ordering, while the monoclinic and trigonal phases do not show any long-range magnetic ordering. Our analysis reveals that the magnetic ordering temperature increases from 170 K (for $x$=0.5) to 280 K (for $x$=1.0) with Ni substitution, accompanied by a simultaneous reduction in magnetization and ordered magnetic moments. Beyond $x \geq 1.25$, neutron diffraction results indicate the absence of long-range magnetic ordering. All compositions exhibit a reentrant spin-glass-like phase below ~50 K, as indicated by magnetization and neutron depolarization data, reflecting intrinsic phase coexistence and magnetic frustration induced by Ni substitution. Furthermore, electrical resistivity measurements reveal a substantial variation in the electric conductivity with increasing Ni content, consistent with a variable-range hopping (VRH) mechanism. This behavior is attributed to the presence of mixed-valence Mn and Ni ions facilitating polaronic hopping. Overall, our results demonstrate the tunable interplay between structure, magnetism, and charge transport in $La_2Mn_{2-x}Ni_xO_6$, establishing its potential for future multifunctional applications, particularly in spintronic devices.

**ACKNOWLEDGMENTS:**

B. Saha acknowledges the financial assistance from Department of Science and Technology (DST), Government of India, for providing the INSPIRE fellowship (Reference No. DST/INSPIRE/03/2017/000817, INSPIRE Grant No. IF180105). SMY acknowledges the financial assistance from ANRF, DST, Govt. of India, under the J. C. Bose fellowship program (JCB/2023/000014).

# REFERENCES:

[1] Y. Markandeya, K. Suresh, and G. Bhikshamaiah, Strong correlation between structural, magnetic and transport properties of non-stoichiometric $Sr_2Fe_xMo_{2-x}O6$ (0.8≤x≤1.5) double perovskites, Journal of Alloys and Compounds **509**, 9598 (2011).

[2] E. C. Ahn, 2D materials for spintronic devices, npj 2D Materials Applications **4**, 17 (2020).

[3] S. Mostufa, S. Liang, V. K. Chugh, J.-P. Wang, and K. Wu, Spintronic devices for biomedical applications, npj Spintronics **2**, 26 (2024).

[4] S. Gardelis, C. Smith, C. Barnes, E. Linfield, and D. Ritchie, Spin-valve effects in a semiconductor field-effect transistor: A spintronic device, Physical Review B **60**, 7764 (1999).

[5] A. K. Bera and S. M. Yusuf, OAJ Materials and Devices, vol 5(1), Chap No 2 in "Perovskites and other framework crystals: new trends and perspectives" (Coll. Acad. 2020) DOI:10.23647/ca.md20202105: Functional Perovskites: Structure-Properties Correlations, OAJ Materials and Devices **5**, 39 (2021).

[6] D. Das and A. Alam, Conical order, magnetic compensation, and sign reversible exchange bias in spinel structured $AB_2O_4$ compounds: A Monte Carlo study, Physical Review Materials **5**, 044404 (2021).

[7] M. Matsuda, H. Ueda, A. Kikkawa, Y. Tanaka, K. Katsumata, Y. Narumi, T. Inami, Y. Ueda, and S.-H. Lee, Spin–lattice instability to a fractional magnetization state in the spinel $HgCr_2O_4$, Nature Physics **3**, 397 (2007).

[8] V. Ponomar, Crystal structures and magnetic properties of spinel ferrites synthesized from natural Fe–Mg–Ca carbonates, Materials Research Bulletin **158**, 112068 (2023).

[9] V. Ivanov, M. Abrashev, M. Iliev, M. Gospodinov, J. Meen, and M. Aroyo, Short-range B-site ordering in the inverse spinel ferrite $NiFe_2O_4$, Physical Review B—Condensed Matter **82**, 024104 (2010).

[10] S. Thota, M. Reehuis, A. Maljuk, A. Hoser, J.-U. Hoffmann, B. Weise, A. Waske, M. Krautz, D. Joshi, and S. Nayak, Neutron diffraction study of the inverse spinels $Co_2TiO_4$ and $Co_2SnO_4$, Physical Review B **96**, 144104 (2017).

[11] S.-D. Mo and W. Ching, Electronic structure of normal, inverse, and partially inverse spinels in the $MgAl_2O_4$ system, Physical Review B **54**, 16555 (1996).

[12] H. Tian, X.-Y. Kuang, A.-J. Mao, Y. Yang, H. Xiang, C. Xu, S. O. Sayedaghaee, J. Íñiguez, and L. Bellaiche, Novel type of ferroelectricity in brownmillerite structures: A first-principles study, Physical Review Materials **2**, 084402 (2018).

[13] H. Tian, L. Bellaiche, and Y. Yang, Diversity of structural phases and resulting control of properties in brownmillerite oxides: A first-principles study, Physical Review B **100**, 220103 (2019).

[14] A. Colville and S. Geller, The crystal structure of brownmillerite, $Ca_2FeAlO_5$, Acta Crystallographica Section B: Structural Crystallography Crystal Chemistry **27**, 2311 (1971).

[15] P. Sarkar, P. Mandal, A. K. Bera, S. M. Yusuf, L. Sharath Chandra, and V. Ganesan, Field-induced first-order to second-order magnetic phase transition in $Sm_{0.52}Sr_{0.48}MnO_3$, Physical Review B **78**, 012415 (2008).

[16] P. Sarkar, P. Mandal, K. Mydeen, A. K. Bera, S. M. Yusuf, S. Arumugam, C. Jin, T. Ishida, and S. Noguchi, Role of external and internal perturbations on the ferromagnetic phase transition in $Sm_{0.52}Sr_{0.48}MnO_3$, Physical Review B **79**, 144431 (2009).

[17] Pooja, K. S. Chikara, Aprajita, S. N. Sarangi, D. Samal, S. Saha, A. K. Bera, S. M. Yusuf, and C. Sow, Stabilization of ferromagnetism via structural modulations in Cr-doped $CaRuO_3$: A neutron diffraction and Raman spectroscopy study, Physical Review B **110**, 184425 (2024).

[18] K. Manna, A. K. Bera, M. Jain, S. Elizabeth, S. M. Yusuf, and P. Anil Kumar, Structural-modulation-driven spin canting and reentrant glassy magnetic phase in ferromagnetic $Lu_2MnNiO_6$, Physical Review B **91**, 224420 (2015).

[19] K. Manna, R. Sarkar, S. Fuchs, Y. Onykiienko, A. K. Bera, G. A. Cansever, S. Kamusella, A. Maljuk, C. Blum, and L. Corredor, Noncollinear antiferromagnetism of coupled spins and pseudospins in the double perovskite $La_2CuIrO_6$, Physical Review B **94**, 144437 (2016).

[20] R. K. Patel, K. S. Chikara, S. M. Hossain, M. Majumder, C. Nath, S. M. Yusuf, M. P. Saravanan, A. K. Bera, and A. K. Pramanik, Crystal structure, magnetic, and magnetostructural correlations in double perovskite antiferromagnets $Sr_2NiMo_{1-}$, Physical Review Materials **9**, 044406 (2025).

[21] K. D. Chandrasekhar, A. Das, C. Mitra, and A. Venimadhav, The extrinsic origin of the magnetodielectric effect in the double perovskite $La_2NiMnO_6$, Journal of Physics: Condensed Matter **24**, 495901 (2012).
[22] R. Borges, R. Thomas, C. Cullinan, J. Coey, R. Suryanarayanan, L. Ben-Dor, L. Pinsard-Gaudart, and A. Revcolevschi, Magnetic properties of the double perovskites $A_2FeMoO_6$ A= Ca, Sr, Ba, Journal of Physics: Condensed Matter **11**, L445 (1999).
[23] M. Azuma, S. Kaimori, and M. Takano, High-pressure synthesis and magnetic properties of layered double perovskites $Ln_2CuMO_6$ (Ln= La, Pr, Nd, and Sm; M= Sn and Zr), Chemistry of materials **10**, 3124 (1998).
[24] T. Goto, T. Kimura, G. Lawes, A. Ramirez, and Y. Tokura, Ferroelectricity and giant magnetocapacitance in perovskite rare-earth manganites, Physical Review Letters **92**, 257201 (2004).
[25] R. Das and R. N. P. Choudhary, Electrical and magnetic properties of double perovskite: $Y_2CoMnO_6$, Ceramics International **47**, 439 (2021).
[26] B. Miao, S. Huang, D. Qu, and C. Chien, Inverse spin Hall effect in a ferromagnetic metal, Physical Review Letters **111**, 066602 (2013).
[27] J. Rodríguez-Carvajal, M. Hennion, F. Moussa, A. H. Moudden, L. Pinsard, and A. Revcolevschi, Neutron-diffraction study of the Jahn-Teller transition in stoichiometric $LaMnO_3$, Physical Review B **57**, R3189 (1998).
[28] C. Ritter, M. Ibarra, L. Morellon, J. Blasco, J. Garcia, and J. De Teresa, Structural and magnetic properties of double perovskites AA'$FeMoO_6$ (AA'= $Ba_2$, BaSr, $Sr_2$ and $Ca_2$), Journal of Physics: Condensed Matter **12**, 8295 (2000).
[29] J. Blasco, J. Garcia, M. Sanchez, A. Larrea, J. Campo, and G. Subias, Magnetic properties and structure of $LaNi_{3/4}Mn_{1/4}O_3$, Journal of Physics: Condensed Matter **13**, L729 (2001).
[30] C. Bull, D. Gleeson, and K. Knight, Determination of B-site ordering and structural transformations in the mixed transition metal perovskites $La_2CoMnO_6$ and $La_2NiMnO_6$, Journal of Physics: Condensed Matter **15**, 4927 (2003).
[31] J. Philipp, P. Majewski, L. Alff, A. Erb, R. Gross, T. Graf, M. Brandt, J. Simon, T. Walther, and W. Mader, Structural and doping effects in the half-metallic double perovskite $A_2CrWO_6$ (A= Sr, Ba, and Ca), Physical Review B **68**, 144431 (2003).
[32] S. Kumar, G. Giovannetti, J. van den Brink, and S. Picozzi, Theoretical prediction of multiferroicity in double perovskite $Y_2NiMnO_6$, Physical Review B **82**, 134429 (2010).
[33] K.-I. Kobayashi, T. Kimura, H. Sawada, K. Terakura, and Y. Tokura, Room-temperature magnetoresistance in an oxide material with an ordered double-perovskite structure, Nature **395**, 677 (1998).
[34] N. Kallel, J. Dhahri, S. Zemni, E. Dhahri, M. Oumezzine, M. Ghedira, and H. Vincent, Effect of Cr Doping in $La_{0.7}Sr_{0.3}Mn_{1-x}Cr_xO_3$ with 0≤ x≤ 0.5, Physica Status Solidi **184**, 319 (2001).
[35] J. Ahmad, M. Iqbal, J. A. Khan, S. H. Bukhari, M. T. Jamil, and U. Nissar, in *First iiScience International Conference 2020* (SPIE, 2020), pp. 69.
[36] C. Gauvin-Ndiaye, A. M. S. Tremblay, and R. Nourafkan, Electronic and magnetic properties of the double perovskites ${\mathrm{La}}_{2}{\mathrm{MnRuO}}_{6}$ and $\mathrm{LaA}{\mathrm{MnFeO}}_{6}\phantom{\rule{4pt}{0ex}}(A=\mathrm{Ba},\text{Sr},\text{Ca})$ and their potential for magnetic refrigeration, Physical Review B **99**, 125110 (2019).
[37] M. Ullah, S. A. Khan, G. Murtaza, R. Khenata, N. Ullah, and S. B. Omran, Electronic, thermoelectric and magnetic properties of $La_2NiMnO_6$ and $La_2CoMnO_6$, Journal of Magnetism Magnetic Materials **377**, 197 (2015).
[38] D. Choudhury, P. Mandal, R. Mathieu, A. Hazarika, S. Rajan, A. Sundaresan, U. Waghmare, R. Knut, O. Karis, and P. Nordblad, Near-room-temperature colossal magnetodielectricity and multiglass properties in partially disordered $La_2NiMnO_6$, Physical Review Letters **108**, 127201 (2012).
[39] A. Biswal, J. Ray, P. Babu, V. Siruguri, and P. Vishwakarma, Dielectric relaxations in $La_2NiMnO_6$ with signatures of Griffiths phase, Journal of Applied Physics **115**, 194106 (2014).
[40] S. C. Baral, P. Maneesha, E. Rini, and S. Sen, Recent advances in $La_2NiMnO_6$ Double Perovskites for various applications; Challenges and opportunities, arXiv preprint arXiv:.06951 (2023).
[41] R. D. Sánchez, M. T. Causa, J. Sereni, M. Vallet-Regí, M. J. Sayagués, and J. M. González-Calbet, Specific heat, magnetic susceptibility and electrical resistivity measurements on $LaNiO_3$, Journal of Alloys and Compounds **191**, 287 (1993).
[42] J. B. Goodenough, Metallic oxides, Progress in Solid State Chemistry **5**, 145 (1971).
[43] M. Nasir, S. Kumar, A. K. Bera, M. Khan, S. Yusuf, and S. Sen, Cluster spin-glass-like dynamics in Mn-rich $La_2Ni_{0.5}Mn_{1.5}O_6$ ferromagnetic insulator, Journal of Physics: Condensed Matter **33**, 225803 (2021).

[44] M. Nasir, A. K. Bera, S. M. Yusuf, N. Patra, D. Bhattacharya, S. Kumar, S. Jha, S.-W. Liu, S. Biring, and S. Sen, Structural, electrical, and magnetic properties of $La_2Ni_{1.5}Mn_{0.5}O_6$ double perovskites, arXiv preprint arXiv:.07091 (2017).
[45] M. Nasir, M. Khan, S. Bhatt, A. K. Bera, M. Furquan, S. Kumar, S. M. Yusuf, N. Patra, D. Bhattacharya, and S. N. Jha, Influence of cation order and valence states on magnetic ordering in $La_2Ni_{1-x}Mn_{1+x}O_6$, Physica Status Solidi **256**, 1900019 (2019).
[46] J. Blasco, J. García, M. Sánchez, J. Campo, G. Subías, and J. Pérez-Cacho, Magnetic properties of $LaNi_{1-x}Mn_xO_{3+\delta}$ perovskites, The European Physical Journal B-Condensed Matter Complex Systems **30**, 469 (2002).
[47] J. Rodríguez-Carvajal, FullProf, CEA/Saclay, France **1045**, 132 (2001).
[48] S. M. Yusuf and L. Madhav Rao, Magnetic studies in mesoscopic length scale using polarized neutron spectrometer at Dhruva reactor, Trombay, Pramana **47**, 171 (1996).
[49] A. Wold, R. J. Arnott, and J. B. Goodenough, Some Magnetic and Crystallographic Properties of the System $LaMn1_{-x}Ni_xO_{3+\lambda}$, Journal of Applied Physics **29**, 387 (1958).
[50] J. B. Goodenough, A. Wold, R. Arnott, and N. Menyuk, Relationship between crystal symmetry and magnetic properties of ionic compounds containing $Mn^{3+}$, Physical Review B **124**, 373 (1961).
[51] R. Dass, J.-Q. Yan, and J. Goodenough, Oxygen stoichiometry, ferromagnetism, and transport properties of $La_{2-x}NiMnO_{6+\delta}$, Physical Review B **68**, 064415 (2003).
[52] F. N. Sayed, S. Achary, O. Jayakumar, S. Deshpande, P. Krishna, S. Chatterjee, P. Ayyub, and A. Tyagi, Role of annealing conditions on the ferromagnetic and dielectric properties of $La_2NiMnO_6$, Journal of Materials Research **26**, 567 (2011).
[53] B. Ravel, E. Stern, R. Vedrinskii, and V. Kraizman, Local structure and the phase transitions of $BaTiO_3$, Ferroelectrics **206**, 407 (1998).
[54] C. Ritter, M. R. Ibarra, J. M. De Teresa, P. A. Algarabel, C. Marquina, J. Blasco, J. García, S. Oseroff, and S. W. Cheong, Influence of oxygen content on the structural, magnetotransport, and magnetic properties of $LaMnO_{3+\delta}$, Physical Review B **56**, 8902 (1997).
[55] P. W. Anderson, Antiferromagnetism. Theory of Superexchange Interaction, Physical Review **79**, 350 (1950).
[56] J. B. Goodenough, Theory of the Role of Covalence in the Perovskite-Type Manganites [La, M(II)]$MnO_3$, Physical Review **100**, 564 (1955).
[57] S. M. Yusuf and L. M. Rao, The magnetic domain effect in the local canted spin ferrite $Zn_{0.5}Co_{0.5}Fe_{2-x}Cr_xO_4$: a macroscopic and mesoscopic study, Journal of Physics: Condensed Matter **7**, 5891 (1995).
[58] A. Kumar, S. Giri, T. Nath, C. Ritter, and S. Yusuf, Investigation of magnetic ordering and origin of exchange-bias effect in doped manganite, $Sm_{0.4}Ca_{0.6}MnO_3$, Journal of Applied Physics **128** (2020).
[59] J. Blasco, M. Sanchez, J. Perez-Cacho, J. Garcıa, G. Subıas, and J. Campo, Synthesis and structural study of $LaNi1_{-x}Mn_xO_{3+\delta}$ perovskites, Journal of Physics Chemistry of Solids **63**, 781 (2002).
[60] M. Nasir, S. Kumar, A. K. Bera, M. Khan, S. M. Yusuf, and S. Sen, Cluster spin-glass-like dynamics in Mn-rich $La_2Ni_{0.5}Mn_{1.5}O_6$ ferromagnetic insulator, Journal of Physics: Condensed Matter **33**, 225803 (2021).
[61] D. K. Mahato, A. Molak, I. Gruszka, A. Winiarski, and J. Koperski, Low temperature - Dielectric, impedance, and conductivity - Study for lanthanum nickelate-manganate ceramics, Physica B: Condensed Matter **640**, 414006 (2022).
[62] D. N. Karmalkar, K. R. Priolkar, K. S. Chikara, A. K. Bera, S. M. Yusuf, and B. Pahari, Enhanced Phase Stability and Oxide-Ion Conductivity in V- and Sr/Ca-Codoped $Bi_2O_3$ Ceramics, The Journal of Physical Chemistry C **129**, 107 (2025).
[63] S. Ivanov, T. Sarkar, G. Bazuev, M. Kuznetsov, P. Nordblad, and R. Mathieu, Modification of the structure and magnetic properties of ceramic $La_2CoMnO_6$ with Ru doping, Journal of Alloys Compounds **752**, 420 (2018).
[64] S. Das, R. Sahoo, S. Shit, and T. Nath, Improved ferromagnetism and transport behaviour in $La_2CoMnO_6$ double perovskite by Ni doping at the Co site, Applied Physics A **128**, 1101 (2022).
[65] J. Shi, Q. Shen, C. Wang, and M. Hu, Magnetic optimization and regulation mechanism of $La_2CoMnO_6$ double-perovskite ceramics by Ti-doping at Mn-site, Ceramics International **49**, 20662 (2023).
[66] C. Agarwal, J. K. Verma, T. Bano, S. Kumawat, M. K. Gora, A. Kumar, S. Chandra, Y. K. Gautam, and S. Kumar, Effect of Cr Substitution for Co on the Structural, Optical, and Magnetic Properties of $La_2CoMnO_6$ Double Perovskite Nanomaterials: A Facile Auto-Combustion Sol-Gel Approach, ECS Journal of Solid State Science Technology **14**, 043012 (2025).

[67] J. Blasco, M. Sánchez, J. Perez-Cacho, J. Garcıa, G. Subıas, and J. Campo, Synthesis and structural study of $LaNi_{1-x}Mn_xO_{3+\delta}$ perovskites, Journal of Physics and Chemistry of Solids **63**, 781 (2002).
[68] S. Das, R. Sahoo, S. Mishra, D. Bhattacharya, and T. Nath, Effects of Ni doping at Co-site on dielectric, impedance spectroscopy and AC-conductivity in $La_2CoMnO_6$ double perovskites, Applied Physics A **128**, 354 (2022).

**Table I:** Rietveld refined crystal structural parameters (lattice parameters, fractional atomic coordinates, site occupancy, and thermal parameters) of $La_2Ni_xMn_{1-x}O_6$ (($x$ = 0.5, 0.75, 1.0, 1.25, and 1.5)) compound at 300 K

| **$x$ = 0.5** | | | | | | |
|---|---|---|---|---|---|---|
| *Space group: Pbnm; a= 5.533(5) Å, b= 5.492(6) Å, c= 7.778(1) Å, α=90°, β=90°, γ =90°* | | | | | | |
| | Wykoff Site | *x*/*a* | *y*/*b* | *z*/*c* | Occupancy | $B_{iso}$ (Å$^2$)×100 |
| La | 4*c* | 1.000 | 0.022(9) | 0.25 | 1.0 | 0.51(1) |
| Ni/Mn | 4*b* | 0.5 | 0 | 0 | 0.25/ 0.75 | 0.62(4) |
| O1 | 4*c* | 0.072(3) | 0.489(1) | 0.25 | 1.0 | 0.62(4) |
| O2 | 8*d* | .724(1) | 0.277(2) | 0.033(6) | 1.0 | 1.21(2) |
| $R_p$ (%)=7.00; $R_{wp}$ (%)=8.98; $R_{exp}$(%)=3.80; $\chi^2$= 5.58 | | | | | | |
| | | | | | | |
| **$x$ = 0.75** | | | | | | |
| *Phase-1 (Orthorhombic)* *Space group: Pbnm; a= 5.5465(1) Å, b=5.4969(9)Å, c=7.8308(2) Å, α=90°, β=90°, γ =90°* | | | | | | |
| | Wyckoff Site | *x*/*a* | *y*/*b* | *z*/*c* | Occupancy | $B_{iso}$ (Å$^2$)×100 |
| La | 4*c* | 0.9621(9) | 0.0704(11) | 0.25 | 1.0 | 0.55(2) |
| Ni/Mn | 4*b* | 0.5 | 0 | 0 | 0.375/0.625 | 0.59(5) |
| O1 | 4*c* | 0.2632(1) | 0.4558(2) | 0.25 | 1.0 | 1.12(3) |
| O2 | 8*d* | 0.5960(1) | 0.2952(7) | 0.0075(8) | 1.0 | 1.12(3) |
| | | | | | | |
| *Phase-2 (Monoclinic)* *Space group: P2₁/c; a= 5.5290(1) Å, b= 5.4871(6) Å, c= 7.7687(9) Å, α=90°, β=89.82(1)°, γ =90°* | | | | | | |
| | Wyckoff Site | *x*/*a* | *y*/*b* | *z*/*c* | Occupancy | $B_{iso}$ (Å$^2$)×100 |
| La | 4*e* | 0.000 | 0.0219(3) | 0.2412(8) | 1.0 | 0.48(2) |
| Ni/Mn | 2*c* | 0 | 0.5 | 0 | 0.56(2)/0.44(2) | 0.65(3) |
| Mn/Ni | 2*d* | 0.5 | 0 | 0 | 0.81(2)/0.19(2) | 0.65(3) |
| O1 | 4*e* | 0.7595(2) | 0.2368(3) | 0.9719(2) | 1.0 | 1.25(5) |
| O2 | 4*e* | 0.2150(3) | 0.7800(5) | 0.4648(2) | 1.0 | 1.25(5) |
| O3 | 4*e* | 0.5679 | 0.9954(1) | 0.2433(2) | 1.0 | 1.25(5) |
| $R_p$ (%)=7.17; $R_{wp}$ (%)=9.57; $R_{exp}$(%)=3.94 ; $\chi^2$= 5.89 | | | | | | |
| | | | | | | |
| **$x$ = 1** | | | | | | |
| *Phase-1 (Orthorhombic)* *Space group: Pbnm; a= 5.4931(1) Å, b=5.4484(6) Å, c=7.7339(3) Å, α=90°, β=90°, γ =90°* | | | | | | |
| | Wyckoff Site | *x*/*a* | *y*/*b* | *z*/*c* | Occupancy | $B_{iso}$ (Å$^2$)×100 |
| La | 4*c* | 0.9576(4) | 0.0316(3) | 0.25 | 1.0 | 0.51(6) |
| Ni/Mn | 4*b* | 0.5 | 0 | 0 | 0.5/0.5 | 0.57(4) |
| O1 | 4*c* | 0.0584(8) | 0.5031(5) | 0.25 | 1.0 | 1.17(2) |
| O2 | 8*d* | 0.7331(1) | 0.2740(7) | 0.0301(3) | 1.0 | 1.17(2) |
| *Phase-2 (Monoclinic)*<br>*Space group: $P2_1/c$; a=5.4619(1) Å, b=5.5095(6) Å, c=7.7778(11) Å, α=90º, β=90.36(1)º, γ =90º* | | | | | | |
| | Wyckoff Site | *x*/*a* | *y*/*b* | *z*/*c* | Occupancy | $B_{iso}$ (Å$^2$)×100 |
| La | 4*e* | 0.0108(7) | 0.0048(3) | 0.2744(3) | 1.0 | 0.56(4) |
| Ni/Mn | 2*c* | 0 | 0.5 | 0 | 0.81(3)/0.19(3) | 0.60(6) |
| Mn/Ni | 2*d* | 0.5 | 0 | 0 | 0.81(3)/0.19(3) | 0.60(6) |
| O1 | 4*e* | 0.2561(5) | 0.2492(4) | 0.05192) | 1.0 | 1.19(2) |
| O2 | 4*e* | 0.2816(5) | 0.3044(2) | 0.4640(5) | 1.0 | 1.19(2) |
| O3 | 4*e* | 0.5223(3) | 0.0038(5) | 0.2490(4) | 1.0 | 1.19(2) |
| $R_p$ (%) = 6.81; $R_{wp}$ (%) = 8.95; $R_{exp}$(%) = 2.91; $\chi^2$ = 9.46 | | | | | | |
| | | | | | | |
| ***x* = 1.25** | | | | | | |
| *Phase-1 (Trigonal)*<br>*Space group: R-3c; a=5.4998(4) Å, b=5.4998(4) Å, c=13.2011(13) Å, α=90º, β=90º, γ =120º* | | | | | | |
| | Wykoff Site | *x*/*a* | *y*/*b* | *z*/*c* | Occupancy | $B_{iso}$ (Å$^2$)×100 |
| La | 6*a* | 0 | 0 | 0.25 | 1.0 | 0.43(4) |
| Ni/Mn | 6*b* | 0 | 0 | 0 | 0.625/0.325 | 0.69(3) |
| O1 | 18*e* | 0.5586(12) | 0 | 0.25 | 1.0 | 1.28(1) |
| *Phase-2 (Monoclinic)*<br>*Space group: $P2_1/c$; a = 5.5122(1) Å, b = 5.4306(3) Å, c = 7.7311 (4) Å,*<br>*α = 90º, β = 89.82(2)º, γ = 90º* | | | | | | |
| | Wyckoff Site | *x*/*a* | *y*/*b* | *z*/*c* | Occupancy | $B_{iso}$ (Å$^2$)×100 |
| La | 4*e* | 0.9852(4) | 0.0062(3) | 0.2394(2) | 1.0 | 0.51(3) |
| Ni/Mn | 2*c* | 0 | 0.5 | 0.0 | 0.76(2)/0.24(2) | 0.67(4) |
| Mn/Ni | 2*d* | 0.5 | 0 | 0 | 0.51(2)/0.49(2) | 0.67(4) |
| O1 | 4*e* | 0.7628(5) | 0.2132(5) | 0.9863(3) | 1.0 | 1.13(5) |
| O2 | 4*e* | 0.2297(3) | 0.7333(4) | 0.4658(3) | 1.0 | 1.13(5) |
| O3 | 4*e* | 0.5485(7) | 0.0218(5) | 0.2553(4) | 1.0 | 1.13(5) |
| $R_p$ (%)=6.63; $R_{wp}$ (%)=8.95; $R_{exp}$(%)=3.37 ; $\chi^2$= 7.04 | | | | | | |
| | | | | | | |
| ***x* = 1.5** | | | | | | |
| *Phase-1 (Trigonal)*<br>*Space group: R-3c; a = 5.4783(5) Å, b = 5.4783(5) Å, c = 13.1657(3) Å,*<br>*α = 90º, β = 90º, γ = 120º* | | | | | | |
| | Wyckoff Site | *x*/*a* | *y*/*b* | *z*/*c* | Occupancy | $B_{iso}$ (Å$^2$)×100 |
| La | 6*a* | 0 | 0 | 0.25 | 1.0 | 0.46(6) |
| Ni/Mn | 6*b* | 0 | 0 | 0 | 0.75/0.25 | 0.69(5) |
| O1 | 18*e* | 0.5526(11) | 0 | 0.25 | 1.0 | 1.23(3) |
| *Phase-2 (Monoclinic)*<br>*Space group: $P2_1/c$; a = 5.5171(1) Å, b = 5.4018(2) Å, c = 7.7079(3) Å,*<br>*α = 90º, β = 89.90(7)º, γ = 90º* | | | | | | |
| | Wyckoff Site | *x*/*a* | *y*/*b* | *z*/*c* | Occupancy | $B_{iso}$ (Å$^2$)×100 |
| La | 4*e* | 0.0021(4) | 0.0171(5) | 0.2524(1) | 1.0 | 0.54(3) |
| Ni/Mn | 2*c* | 0 | 0 | 0 | 0.81(1)/0.19(1) | 0.62(4) |
| Mn/Ni | 2*d* | 0.5 | 0 | 0 | 0.31(1)/0.69(1) | 0.62(4) |
| O1 | 4*e* | 0.7573(6) | 0.2906(8) | 0.9957(1) | 1.0 | 1.18(1) |
| O2 | 4*e* | 0.1941(7) | 0.8065(5) | 0.4876(9) | 1.0 | 1.18(1) |
| O3 | 4*e* | 0.5696(6) | 0.0114(4) | 0.2573(2) | 1.0 | 1.18(1) |
| $R_p$ (%) = 9.11; $R_{wp}$ (%)=12.2; $R_{exp}$(%)=5.14; $\chi^2$= 5.59 | | | | | | |

**Table II:** Fitted parameters and the agreement factor (adjusted $R^2$) for conductivity data of $La_2Ni_xMn_{1-x}O_6$ ($x$ = 0.5, 0.75, and 1.0) compounds.

| $x$ | Variable Range Hopping Model | | Arrhenius Behaviours | |
|---|---|---|---|---|
| | $E_{dc}$ | $R^2$ | $T_0$ | $R^2$ |
| 0.5 | 1.213(2) | 0.99981 | 1.1198373(10) ×10$^8$ | 0.99993 |
| 0.75 | 1.182(3) | 0.99974 | 6.0874151(2) ×10$^7$ | 0.99766 |
| 1.0 | 1.259(3) | 0.99967 | 2.5068190(2) ×10$^8$ | 0.99949 |

# FIGURES

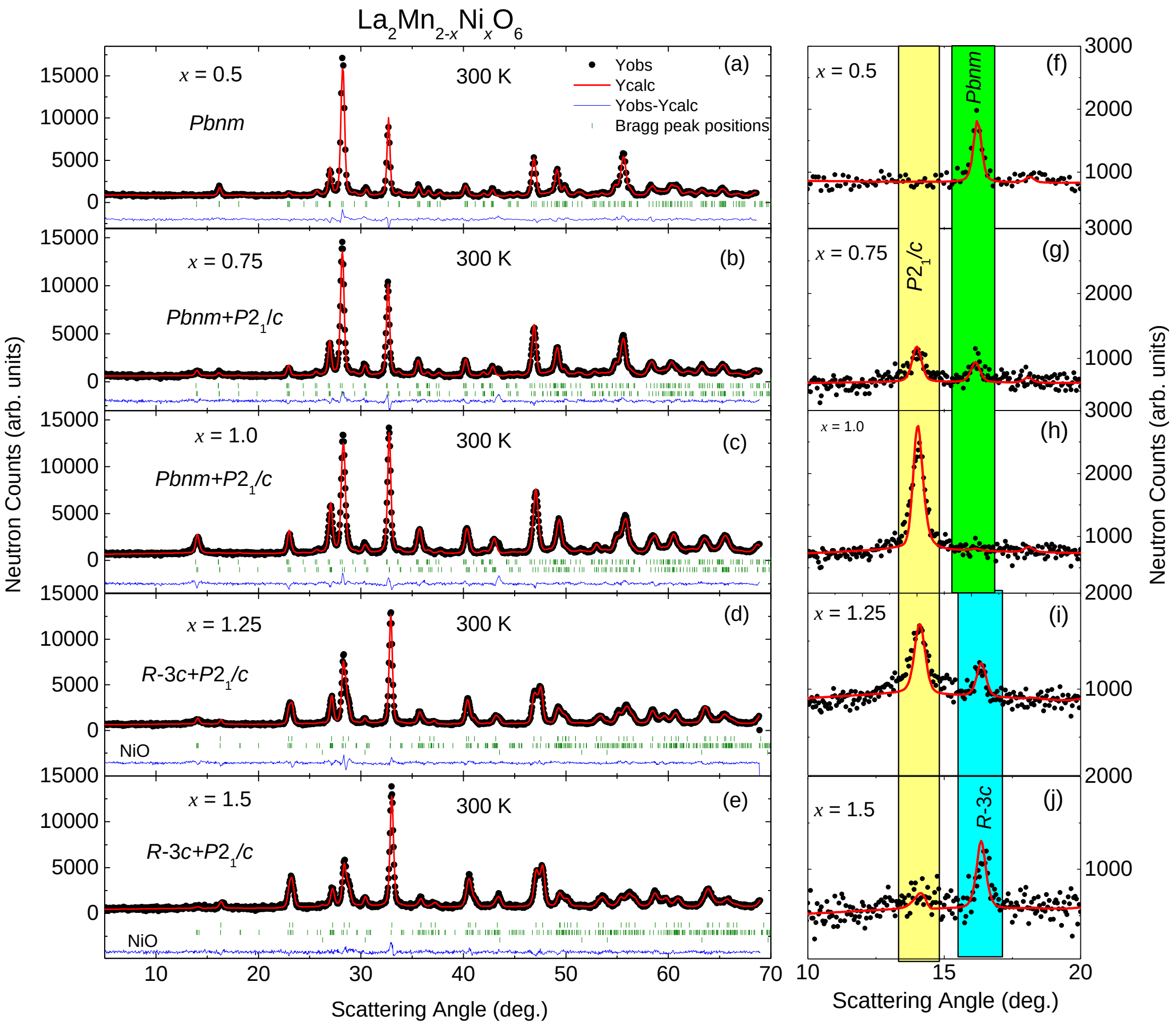

**FIG. 1:** *(a-e) The room temperature neutron powder diffraction patterns for $La_2Mn_{2-x}Ni_xO_6$ (x = 0.5, 0.75, 1.0, 1.25 and 1.5). Experimentally observed and calculated patterns are shown by black circles and solid red lines through the data points. The difference between observed and calculated patterns is shown by the solid lines at the bottom of each panel. The vertical bars indicate the positions of allowed nuclear Bragg peaks (third row in (d) and (e) indicates the positions of allowed nuclear Bragg peaks from the impurity NiO phase). (f-j) The selected area diffraction patterns highlighting the characteristic Bragg peaks for different crystal structures ( yellow: orthorohmbic; green: monoclinic; cyan: trigonal) and the variation of their intensities with Ni-concentrations in $La_2Mn_{2-x}Ni_xO_6$.*

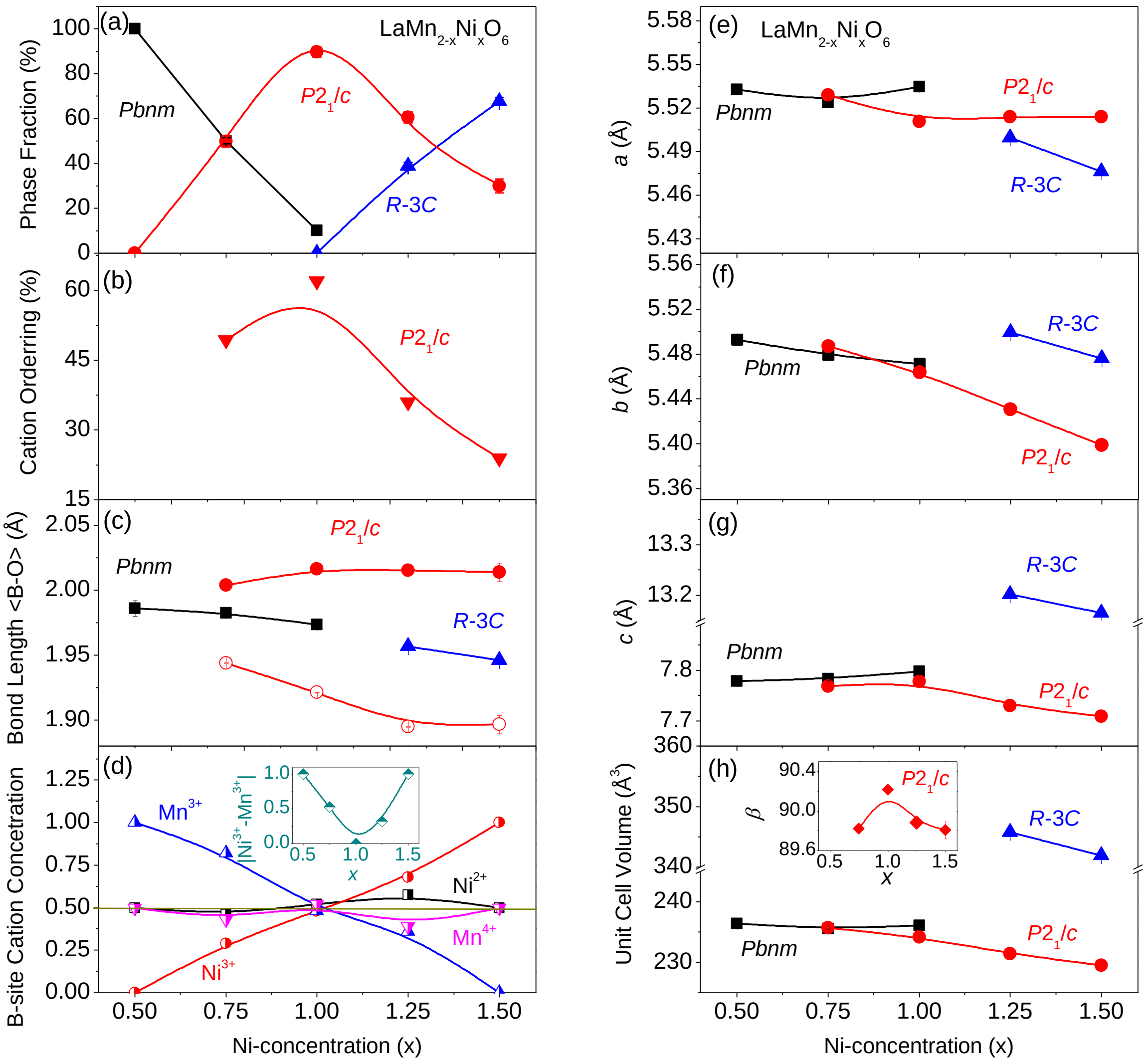

**FIG. 2:** *The Ni concentration (x) in $La_2Mn_{2-x}Ni_xO_6$ (x = 0.5, 0.75, 1.0, 1.25 and 1.5) dependent variations of the (a) fractions of the orthorhombic, monoclinic and triagonal phases, (b) degree of cation ordering at the B and B'sites for the monoclinic phase, (c) average B-O bond-lengths for the B and B' sites, (d) cations ($Mn^{3+}$, $Mn^{4+}$, $Ni^{3+}$, and $Ni^{2+}$) concentrations, (e-g) lattice constants, (h) unit cell volume (V), as derived from the room temperature neutron powder diffraction patterns. The inset of (d) shows the variation of the absolute differences between the concentrations of $Mn^{3+}$ and $Ni^{3+}$ ions with the Ni concentration (x). The inset of (h) shows the variation of the monoclinic angle $\beta$ with the Ni concentration (x). The error bars of the parameters are smaller than the symbol sizes.*

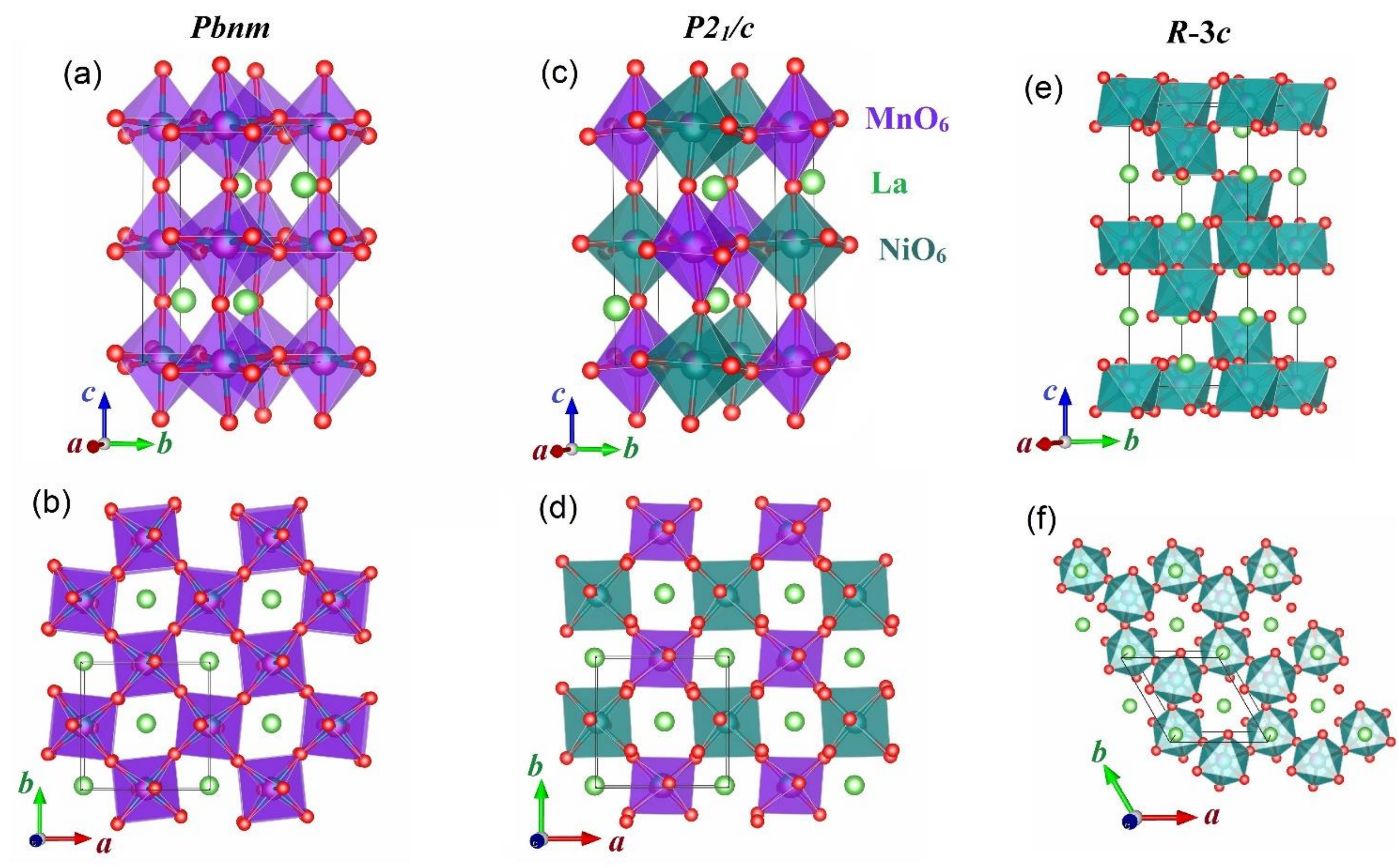

**FIG. 3:** *The schematic crystal structures for the (a-b) Orthorhombic, (c-d) Monoclinic, and (e-f) Trigonal phases of* $La_2Mn_{2-x}Ni_xO_6$.

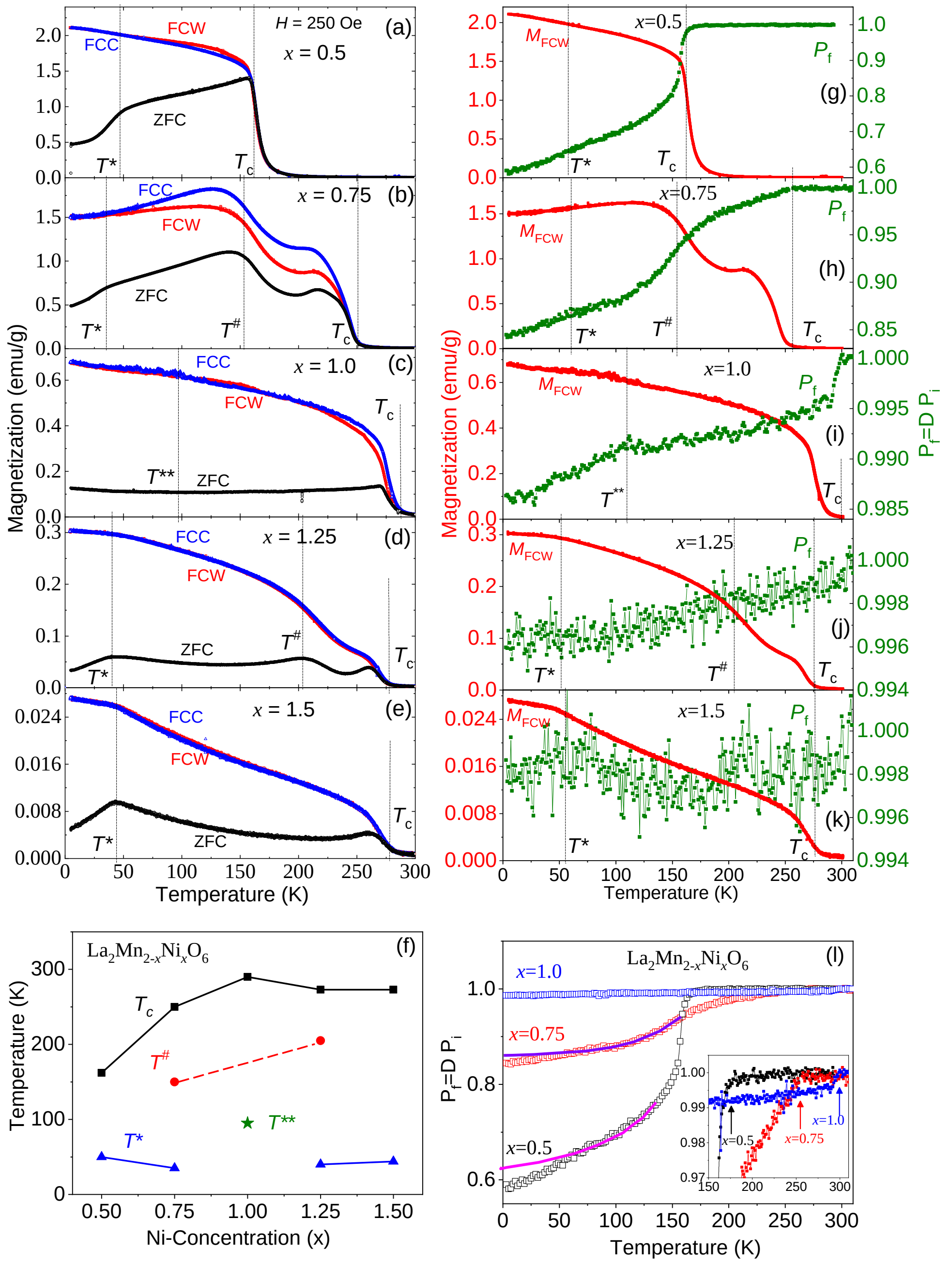

**FIG. 4:** *(a-e) The temperature dependent dc-magnetization curves for $La_2Mn_{2-x}Ni_xO_6$ (x = 0.5, 0.75, 1.0, 1.25 and 1.5), measured under 250 Oe magnetic field in the ZFC, FCC and FCW measurement protocols. (f) The variations of the magnetic transition temperatures $T_c$, $T^{\#}$ and $T^*$ with the Ni-concentration (x). (g-k) The temperature dependent transmitted neutron polarization for $La_2Mn_{2-x}Ni_xO_6$ (x = 0.5, 0.75, 1.0, 1.25 and 1.5) measured under 50 Oe magnetic field under FCW measurement protocol. The FCW magnetization curves are also plotted alongside the depolarization curves for comparison. (l) The over plot of the transmitted neutron polarization curves for x = 0.5, 0.75 and 1.0 compounds to reveal the decrease of the domain magnetization. To reveal the additional contribution at lower temperatures, solid guided lines are shown for the curves corresponding to the x= 0.5 and 0.75 compounds that are the expected behaviour of the depolarization from the major FM phase. Inset shows the zoomed transmitted neutron beam polarization curves highlighting the FM variation of ordering temperature ($T_c$).*

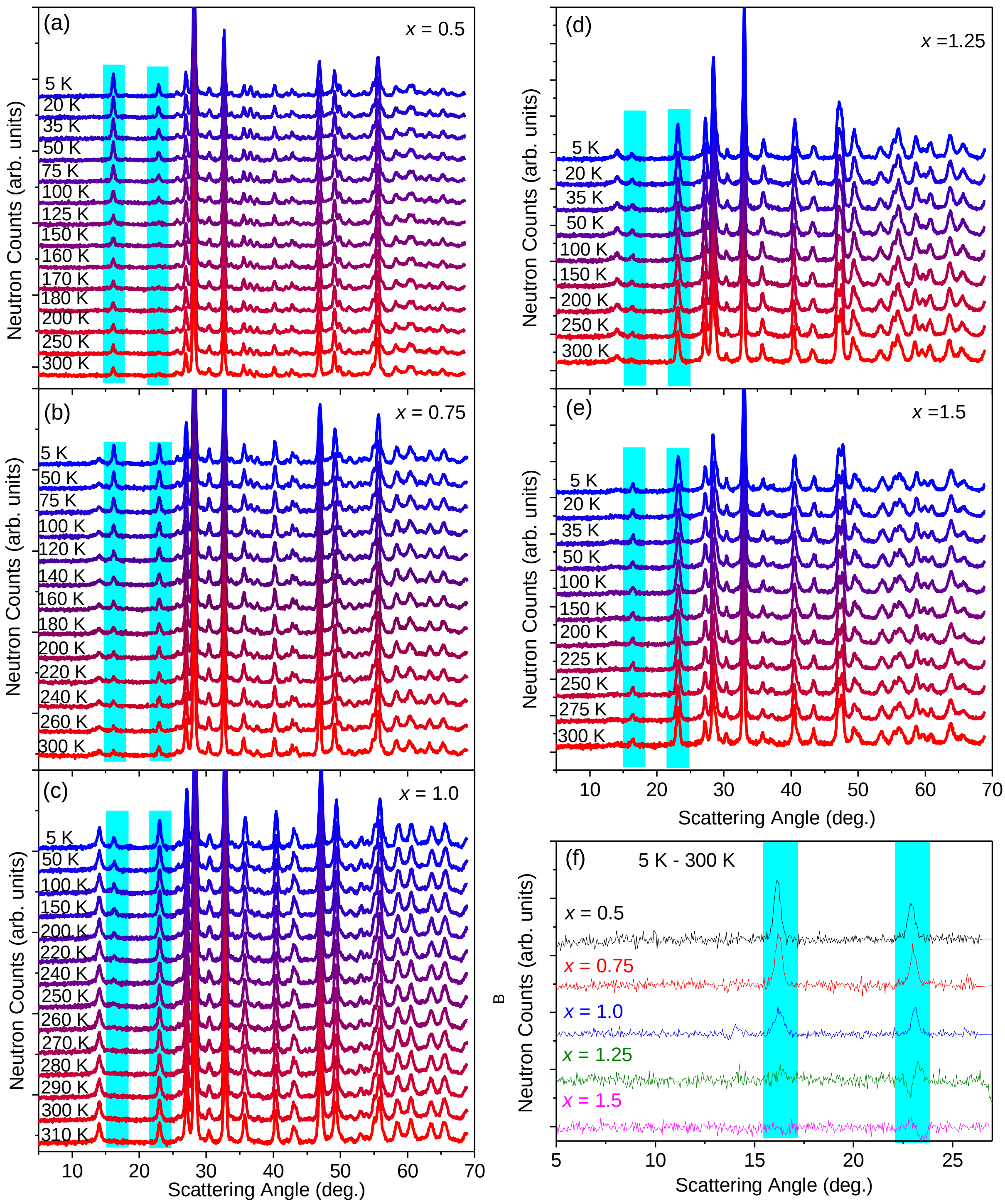

**FIG. 5:** *(a-e) The temperature dependent neutron diffraction patterns for* $La_2Mn_{2-x}Ni_xO_6$ *(x = 0.5, 0.75, 1.0, 1.25 and 1.5). (f) The pure magnetic diffraction patterns at 5 K for* $La_2Mn_{2-x}Ni_xO_6$ *(x = 0.5, 0.75, 1.0, 1.25 and 1.5) obtained after the subtraction of the paramagnetic background measured at 300 K.*

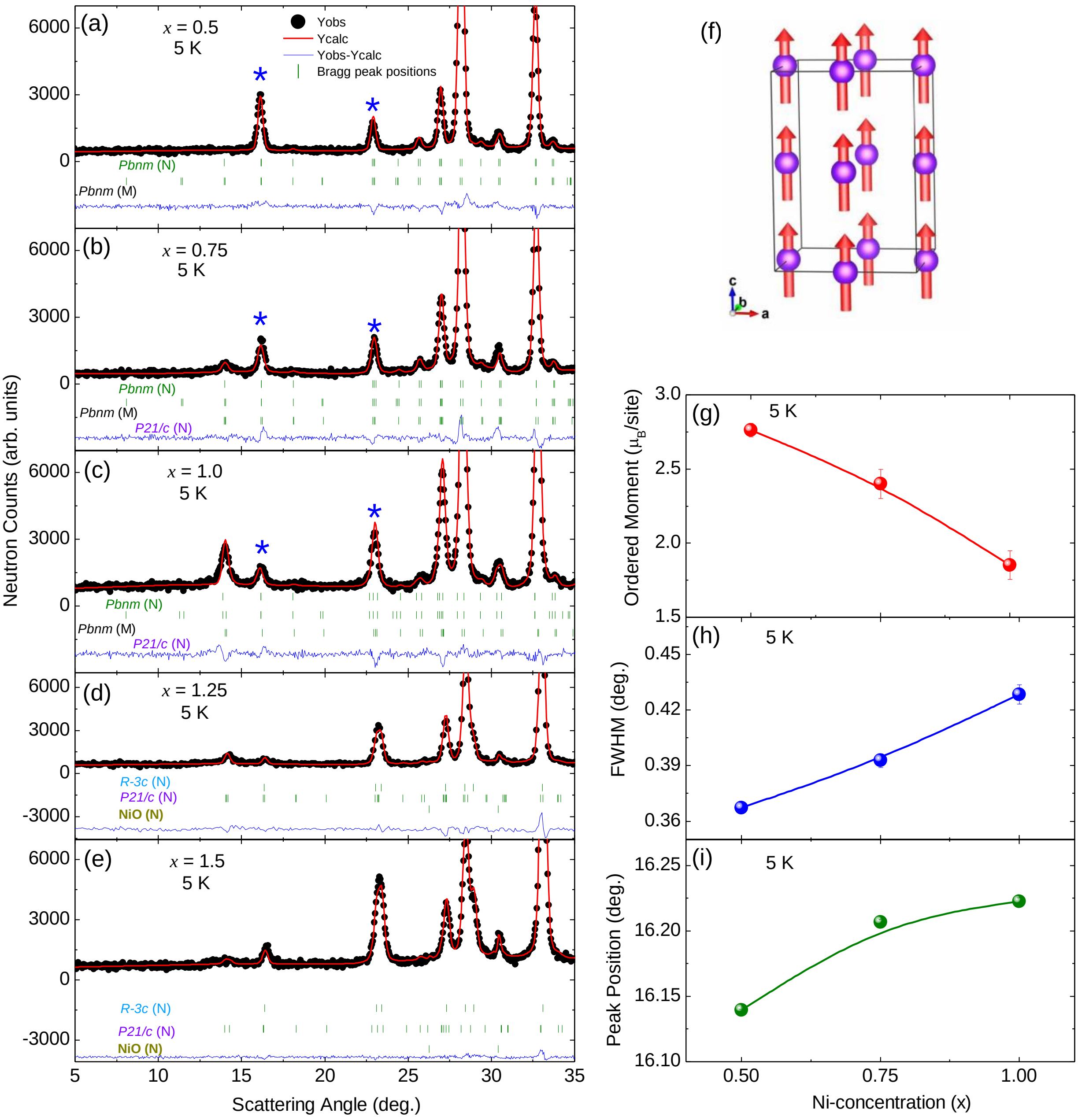

**FIG. 6:** *(a-e) The Rietveld analysed neutron diffraction patterns for* $La_2Mn_{2-x}Ni_xO_6$ *(x = 0.5, 0.75, 1.0, 1.25 and 1.5) at 5 K. The patterns are zoomed over the lower angular and intensity range to highlight the magnetic Bragg peaks at 16 and 23 deg (marked with asterisks). (f) A schematic representation of magnetic structure for* $La_2Mn_{2-x}Ni_xO_6$ *(x = 0.5, 0.75, 1.0, 1.25 and 1.5). (g-i) The variations of ordered site moment, FWHM of the magnetic peak at ~ 16 deg., and peak position, respectively, as a function of Ni-concentration (x).*

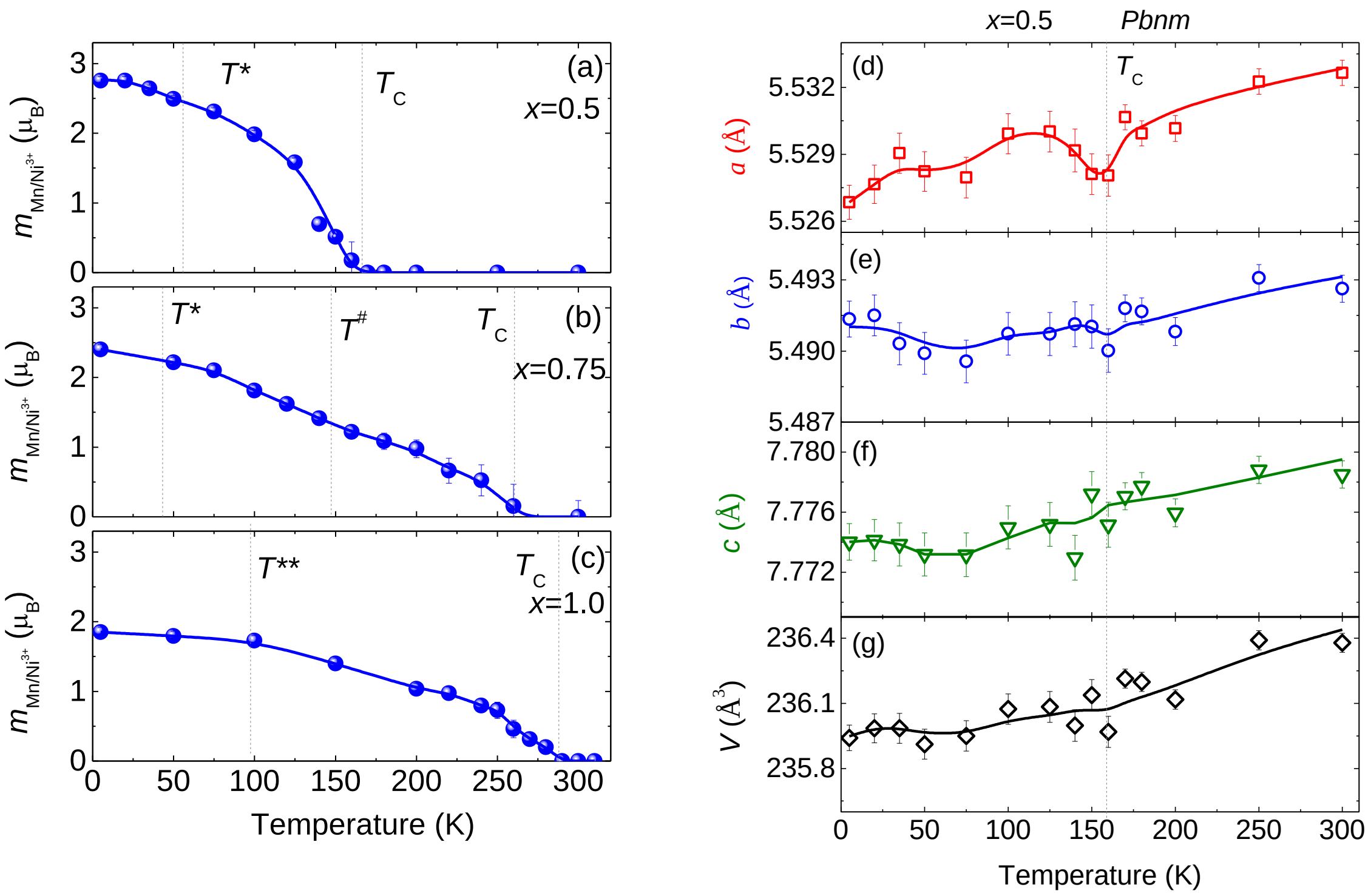

**FIG. 7:** *(a-c) The temperature variations of ordered site moment for* $La_2Mn_{2-x}Ni_xO_6$ *(x = 0.5, 0.75, and 1.0) compounds, respectively. (d-g) The temperature variations of the lattice constants and unit cell volume, respectively. The characteristic temperatures for different magnetic phases are shown by the vertical dashed lines.*

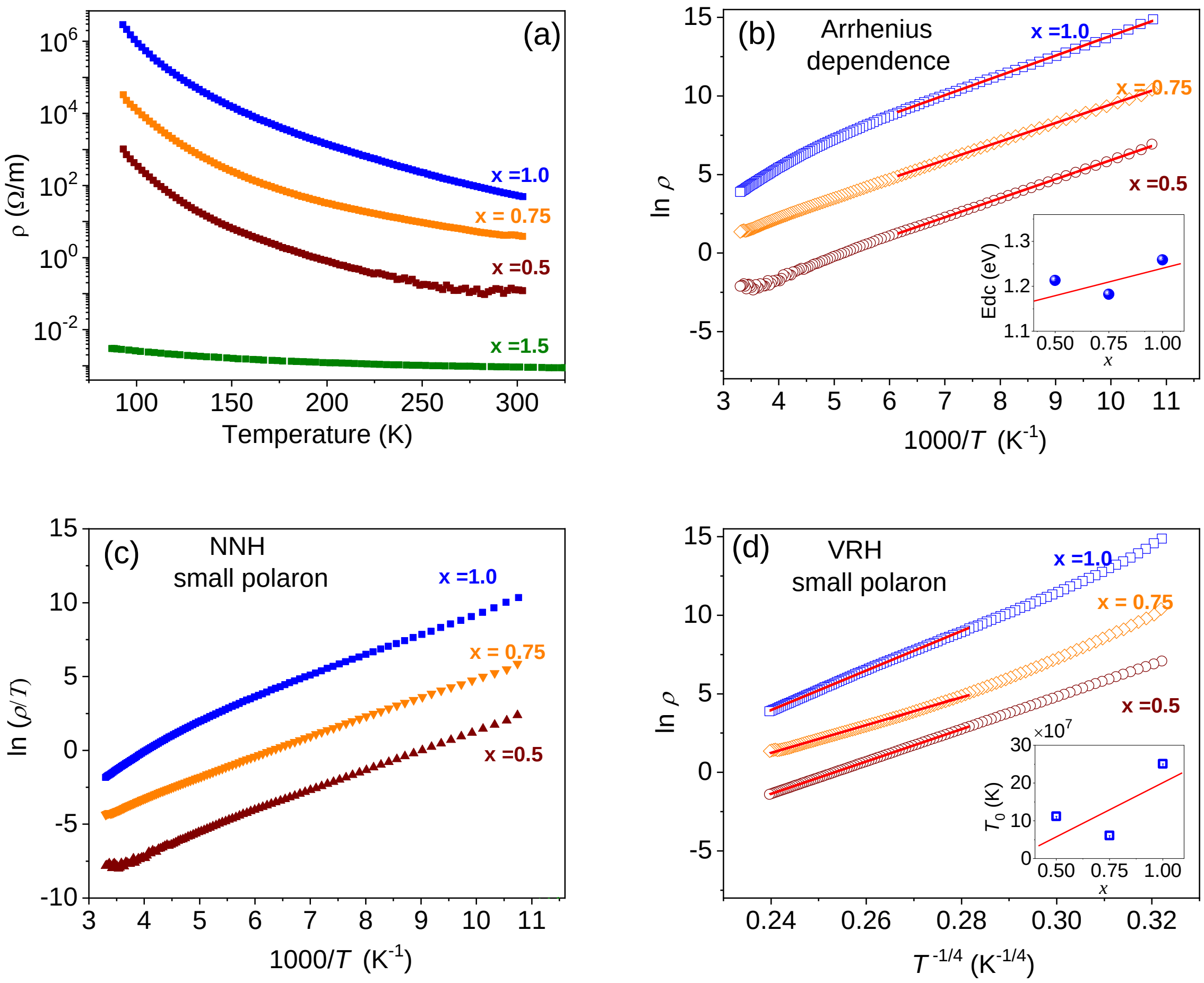

**FIG. 8:** *(a) The temperature dependent resistivity curves for the $La_2Mn_{2-x}Ni_xO_6$ (x = 0.5, 0.75, 1.0, and 1.5) compounds. The data for the x =1.5 compound has been taken from the literature [44]. (b) The logarithmic of resistivity ρ vs 1000/T plots (solid symbol) for the $La_2Mn_{2-x}Ni_xO_6$ (x = 0.5, 0.75, and 1.0) compounds. The lines passing through the data point are the fitted curves by considering Arrhenius model. (c) The logarithmic of resistivity (ρ/T) vs 1000/T plots (solid symbol) for the $La_2Mn_{2-x}Ni_xO_6$ (x = 0.5, 0.75, and 1.0) compounds corresponding to the nearest neighbor hopping (NNH) small polaron model. The lines passing through the data point are the fitted curves by considering VRH model. (d) The logarithmic of resistivity ρ vs $T^{-1/4}$ plots (solid symbol) for the $La_2Mn_{2-x}Ni_xO_6$ (x = 0.5, 0.75, and 1.0) compounds. The lines passing through the data point are the fitted curves by considering VRH model.*